\documentclass[preprint,pre,showpacs,amsmath,amssymb,floatfix]{revtex4} \usepackage{graphicx}

\begin{document}

\title{Resonant nucleation of spatio-temporal order via parametric modal amplification}

\author{Marcelo Gleiser} \email{mgleiser@dartmouth.edu}

\author{Rafael C. Howell\footnote{Present address: Materials Science and Technology Division,
Los Alamos National Laboratory,  Los Alamos, NM 87545, USA}} \email{rhowell@lanl.gov} 

\affiliation{Department of Physics and Astronomy, Dartmouth College, Hanover, NH 03755, USA}

\date{\today}

\begin{abstract} We investigate, analytically and numerically, the emergence of spatio-temporal
order in nonequilibrium scalar field theories. The onset of order is triggered by destabilizing
interactions (DIs), which  instantaneously change the interacting potential from a single to a
double-well, tunable to be either degenerate (SDW) or nondegenerate (ADW). For the SDW case, we
observe the  emergence of spatio-temporal coherent structures known as oscillons. We show that
this emergence is  initially synchronized, the result of parametric amplification of the
relevant oscillon modes. We also discuss how these ordered structures act as bottlenecks for
equipartition. For ADW potentials, we show how the same parametric amplification mechanism may
trigger the rapid decay of  a metastable state. For a range of temperatures, the decay rates
associated with this resonant nucleation can be orders of magnitude larger than those computed
by homogeneous nucleation, with time-scales given by a simple power law, $\tau_{\rm RN}\sim
[E_b/k_BT]^B$, where $B$ depends weakly on the temperature and $E_b/k_BT$ is the free-energy
barrier of a critical fluctuation. \end{abstract}


\maketitle

\section{Introduction} The emergence of spatio-temporal ordered structures in nonlinear systems
is an ideal laboratory for investigating the trend toward complexification observed in nature at
the physical, chemical, and biological level \cite{walgraef}. It is known that localized, or
spatially-bound, order emerges in systems which are out of thermodynamic equilibrium, at the
expense of growing overall disorder \cite{bar-yam}. Examples can be found in hydrodynamics, in
networks of  chemical reactions \cite{cross}, in phase transitions and critical phenomena
\cite{phase,magnets,degennes}, and in living organisms \cite{schnerb}. For these localized 
ordered structures to survive, they must interact with an external environment, which maintains
nonequilibrium conditions: in general, an isolated nonlinear system will eventually reach
equilibrium, maximizing its entropy to the detriment of localized ordered structures.

In other words, if localized ordered structures emerge during the approach to equilibrium of a
closed system,  they will eventually disappear as energy becomes increasingly equipartitioned
amongst the system's many degrees of freedom.  The emergence of localized ordered structures is
thus a nonequilibrium phenomenon, usually corresponding to an unequal partitioning of energy: if
they are long-lived, they will serve as bottlenecks to equipartition. [An exception occurs for
1d field theories with SDW, where configurations such as kink-antikink pairs correspond to the
equilibrium.] Following the pioneering work of Fermi, Pasta, and Ulam  in the early 1950s
\cite{ford}, several recent works have investigated the possibility that discrete kink-like
structures may act as bottlenecks for equipartition in chains of nonlinear oscillators
\cite{discrete,bottlenecks}. 

In field theory and cosmology, most of the interest on ordered configurations has focused on
topological (kinks, vortices, monopoles) \cite{rajaraman,vilenkin} or nontopological
(nontopological solitons, Q-balls) \cite{NTSs} {\it static} solutions of the equations of
motion. Those configurations can be boosted towards each other with a certain velocity and made
to interact, to model nonlinear composite particle interactions. A well-known example are
1-dimensional breathers, time-dependent bound states that have been shown to result from
interactions of kink-antikink pairs \cite{campbell}. However, this is not surprising: any two or
more static solitonic configurations with short-range attractive interactions can be prepared to
generate spatially-bound time-dependent structures \cite{digal}. This is to be contrasted with
the spatially-bound, time-dependent structures  of interest here, that emerge {\it
spontaneously} during the nonlinear evolution of a system. We have recently found such
structures in 2+1-dimensional simulations of nonlinear scalar field models
\cite{gleiser-howell}. Such structures were identified with oscillons, long-lived {\it
time-dependent} localized field configurations that have been previously found deterministically
in a variety of physical systems, ranging from  field theory models with amplitude-dependent
nonlinearities \cite{gleiser,bettinson}, vertically  vibrated grains \cite{umbanhowar}, and
stellar interiors \cite{umurhan}. 

In this paper, we will examine the emergence of spatially-bound, time-dependent field
configurations during  the approach to equilibrium  of (2+1)-dimensional scalar field models
with typical double-well interactions. These interactions are controlled by a tunable external
parameter, which can be interpreted as an extra interaction that may be switched on or off at
will. In the parlance of Ising magnetic systems, the external interaction may be a
spatially-homogeneous magnetic field, ${\cal H}(t)$.  We will investigate both the symmetric
double-well (SDW) -- extending previous results of ref. \cite{gleiser-howell} --  as well as
provide new results for the asymmetric case (ADW). The interest in examining both cases stems 
from the fact that the ensuing dynamics will be quite different: while for the SDW the
degeneracy in free energy density dictates that no critical nucleation is possible, this will
not be the case for the ADW. In fact, for the SDW we observe the synchronous emergence of
long-lived localized configurations (oscillons), which eventually disappear as the system
reaches equipartition \cite{gleiser-howell}. We will argue that these configurations act as
bottlenecks for equipartition, concentrating the field's energy  in a narrow band of
long-wavelength modes. For the ADW, and with the system initially quenched to the metastable
state, the synchronous emergence of oscillons may trigger the nucleation of a critical droplet
of the lower free-energy phase, which grows to complete the phase transformation. We will show
that, for a range of temperatures, the time-scales associated with this process are orders of
magnitude faster than the typical statistical homogeneous nucleation, where a system {\it
adiabatically} cooled into its metastable phase decays to its lowest free energy phase with
typically exponentially-large time scales \cite{langer, domb}. In fact, the present treatment,
with an ``instantaneous'' quench (faster than other typical time-scales in the system),  and the
arbitrarily slow cooling implicit of homogeneous nucleation calculations, represent the two
extreme cases of a whole spectrum of possible cooling rates. Since we will argue that the
emergence of coherent structures is due to the resonant amplification of parametric excitations
of certain field modes, we will refer to this nucleation mechanism as {\it resonant nucleation.}
Anticipating one of our results, for field models described by amplitude-dependent scalar
nonlinearities, resonant nucleation of spatio-temporal order will be relevant whenever the
cooling time scale $\tau_{\rm cool}$ is faster than the equilibration time-scale of the longest
wavelength in the system, $\tau_{\rm eq}^0$.  [The superscript ``0'' is a reminder that often
the longest wavelength is associated with the zero mode of the system, which has the slowest
relaxation time-scale.] We note that we will always work far from the critical point where, of
course, $\tau_{\rm eq}^0\rightarrow \infty$. For our recent work on the approach to criticality
please see ref. \cite{howell}.

The paper is organized as follows. In the next section we discuss the general properties of the
model (a general Ginzburg-Landau model), the initial conditions, and the numerical
implementation of the dynamics. In section 3 we present a mean field (homogeneous Hartree)
approximation adequate to describe the system in the presence of small fluctuations about
equilibrium. The fact that the homogeneous Hartree approximation breaks down as nonperturbative
excitations become progressively more important will provide us with useful information about
the emergence of ordered structures.  In section 4 we argue that the emergence of
spatio-temporal order is a consequence of parametric resonance, induced by the nonlinear
coupling between the field's zero mode -- the forcing agent -- and smaller-$k$ modes. We also
provide  entropic arguments in support of the statement that the emergence of coherent
spatio-temporal structures delays equipartition of the energy and, thus, the approach to
equilibrium. In section 5, we extend our investigation to ADWs. For a range of temperatures [or
initial energies, see below], we measure  the time scales for the decay of the metastable state
as a function of barrier height, contrasting them with those computed from homogeneous
nucleation theory. We identify two possible mechanisms for the decay:  for small asymmetries, a
critical nucleus is produced by the coalescence of two or more oscillons, while for larger
asymmetries a single oscillon evolves into a critical nucleus. We show that, for a range of
temperatures associated with the presence of oscillons in the system,  the time scale associated
with the decay obeys a simple power law, $\tau_{\rm RN}\sim [E_b/k_BT]^B$, where $E_b/k_BT$ is
the free-energy barrier of a critical fluctuation, and the exponent $B$ satisfying $\simeq 2.44
< B(T) < 3.36$.  We conclude in section 6 with a summary of our results and outlook for future
work.

\section{The nonlinear model and its implementation} We start by introducing the model we will
work with and its statistical and numerical implementation. We consider a (2+1)-dimensional real
scalar field (or scalar order parameter) $\phi({\bf x},t)$ evolving under the influence of an
on-site potential $V(\phi)$. If the system does not interact with an external bath, the
continuum Hamiltonian is conserved and gives the total energy of the system, for a given field
configuration, \begin{equation} H[\phi]=\int
d\,^2x\left[\frac{1}{2}(\partial_{t}\phi)^2+\frac{1}{2}(\nabla\phi)^2+V(\phi)\right], \label{H}
\end{equation} where \begin{equation}
V(\phi)=\frac{m^2}{2}\phi^2-\frac{\alpha}{3}\phi^3+\frac{\lambda}{8}\phi^4 \label{V}
\end{equation} is the potential energy density. The parameters $m$,$\alpha$, and $\lambda$ are
positive definite and temperature independent. Before we go any further, a note on our choice of
potential. One could as easily have chosen a typical Ginzburg-Landau (GL)  potential, $V(\phi) =
-\frac{a}{2}\phi^2 + \frac{\lambda}{8}\phi^4 - {\cal H}(t)\phi$, with ${\cal H}(t)$ an external
homogeneous magnetic field. The GL potential can be transformed into the potential of eq.
\ref{V} by a field shift, $\phi' = \phi + C({\cal H};a, \lambda)$, where $C$ is the solution of
the cubic equation $C^3-2\frac{a}{\lambda} C+  \frac{2H}{\lambda}=0$, $\alpha =
\frac{3}{2}\lambda C$, and $m^2=-a + \frac{3}{2}C^2\lambda$. We chose the potential of eq.
\ref{V} for convenience. Our control parameter is $\alpha$. We note that when $\alpha=0$ the
potential is an anharmonic single well. In this case, there are no spinodal instabilities. (By
spinodal instabilities we mean $V''(\phi)<0$ for certain values of $\phi$, which may result in
the unstable growth of modes.)  When $\alpha \geq 3/2$ (in scaled units, see below) the
potential can assume a typical double-well shape [SDW for equality and ADW for $\alpha > 3/2$], 
and spinodal instabilities are possible. We will thus refer to the interactions modeled by
$\alpha$ as destabilizing interactions (DIs). Changing $\alpha$ from an initial value of zero to
a finite positive value can also be thought of as an ``instantaneous'' quench: the system,
initially prepared in a single well, is ``tossed'' into a double-well potential within a
time-scale much faster than any equilibration time-scale for longer wavelengths in the system,
those with $k^2 \lesssim 2\pi V''(\phi=0)$, where $k$ is the mode's wave number. We will say
more about this later on.

It is helpful to introduce the dimensionless variables $\phi'=\phi\,\sqrt\lambda/m$, $x'=xm$,
$t'=tm$, and $\alpha'=\alpha/(m\sqrt\lambda)$ (henceforth dropping the primes).  When $\alpha
\equiv \alpha_c= 3/2$ the two minima are degenerate (SDW). If $\alpha \neq 0$, the ${\cal Z}_2$
symmetry is explicitly broken. We will only be concerned here with the cases modeled by $\alpha
\geq \alpha_c$. For $\alpha > \alpha_c$ the minimum at $\phi=0$ becomes the local (or
metastable) minimum (ADW) and  $\phi_+= \alpha + \sqrt{\alpha^2 - 2}$ is the global minimum. 

\subsection{Statistical Implementation and Initial Conditions} A statistical description of the
state of the system is given by a set of time-dependent  observables  $\{\langle
O^{(1)}(t)\rangle,\langle O^{(2)}(t)\rangle,\ldots\}$, each one calculated  as an ensemble
average over $N$ realizations, or microstates. For example, \begin{equation} \langle
O^{(1)}(t)\rangle=N^{-1}\sum_{i=1}^{N}O^{(1)}_i(t). \end{equation} In most cases, an observable
within a particular microstate is itself an average over the area of the system, such as the
average value of the field  $\phi_i(t)=A^{-1}\int_{A}\!d\,^2x\,\phi_i({\bf x},t)$ for  the
particular microstate $i$. Most of the observables considered will have this ``double''
averaging. Such is the case for $\langle\phi(t)\rangle$, $\langle\pi(t)\rangle$ (the field's
conjugate momentum), and the corresponding hierarchy of correlation functions.  It will be noted
when this is not the case, particularly when quantities are no longer assumed to be
translationally invariant, such as those describing localized configurations. 

We prepare the field such that it is in thermal equilibrium at temperature $T$ in the
single-well ($\alpha=0$) potential at some time prior to the quench at $t=0$, when $\alpha$ is
set to a nonzero value. When one is faced with a nonlinear equation of motion such as eq.
\ref{lang} below, the thermal probability distribution for $\phi({\bf x},0)$ cannot be obtained
exactly.  To generate the initial conditions for each of the microstates in the ensemble, we
instead couple the field in the single-well potential to a heat  bath via a generalized Langevin
equation with additive noise, \begin{equation}
\ddot\phi+\gamma\dot\phi-\nabla^2\phi=-\phi-\frac{1}{2}\phi^3+\xi, \label{lang} \end{equation}
where the viscosity coefficient $\gamma$ is related to the stochastic force of zero mean
$\xi({\bf x},t)$ by the fluctuation-dissipation relation, \begin{equation} \langle \xi({\bf
x},t)\,\xi({\bf x'},t')\rangle=2\gamma T\,\delta^2({\bf x}-{\bf x'})\,\delta(t-t'). \label{fd}
\end{equation} Eq. \ref{fd} defines the two-point correlation function of the stochastic force,
assumed here to be Markovian. Since this force will only be used during the preparation of the
initial thermal state -- after the quench the coupling with the bath is turned off ($\gamma
\rightarrow 0$) -- this assumption is sufficient. [However, it would be quite interesting to
examine the effects of multiplicative noise in the nonequilibrium dynamics of this system when
the contact with the thermal bath is kept throughout its evolution. [The role of multiplicative
noise on the onset of synchronization in 1+1-dimensional fields has been recently investigated
by Mun\~oz and Pastor-Satorras \cite{munoz}.] We note that even though we are preparing each
microstate in an initial thermal state, there is a one to one correspondence between the
temperature and the average energy of the field. In fact, we can refer to the ensemble prepared
as above as a microcanonical ensemble, since each microstate has energy $E \pm \delta E$, where
$\delta E/E <  10^{-3}$. Such numerical accuracy in energy matches that obtained with other
methods used preparing the initial state, including the popular Monte Carlo sampling.  It simply
expresses the fact that, for large enough lattices that approach the thermodynamic limit, there
is an equivalence between canonical and microcanonical ensembles \cite{chandler}.

\subsection{Numerical Implementation}

The full evolution of the scalar field for each microstate is simulated using a staggered 
leapfrog integration of the equation of motion. It is implemented on a square lattice with
periodic boundary conditions.  Integrations are performed with $\Delta x=0.2$ and $\Delta
t=0.02$ with 1024 lattice sites per side. The field is then defined over an area with sides of
length $L=204.8$. The  leapfrog scheme replaces second-order derivatives with  finite-difference
expressions that are accurate to ${\cal O}(\Delta x^2,\Delta t^2)$. The lattice spacing and
system size are chosen so that computations can be carried out in a timely  manner, while the
numerical values of any observables and their corresponding continuum values are  in good
agreement. For example, after the quench, the energy in this closed system  is conserved by more
than 1 part in $10^3$. Results are independent of lattice size, provided large enough lattices
are used.  Finally, all of the observables are then averaged over an ensemble  of $N=100$
microstates.

\section{Dynamics within the perturbative realm: Homogeneous Hartree Approximation}

\subsection{Setting up the initial configuration} We start our investigation using a familiar
perturbative method, the leading order or Homogeneous Hartree Approximation (HHA). An excellent
account can be found in \cite{bonini}, where  successive truncations within the hierarchy of
correlation functions were used to simplify the equation of motion, all the while capturing the
essential aspects of the dynamics.  We will borrow many of the same methods and notation found
in this reference.  The HHA assumes that the statistical distribution of the fluctuations in the
field and its momentum remain Gaussian throughout the evolution of the system.  We will see that
this is indeed an excellent approximation in the following cases: i) just prior to the quench
(that is, the initial state of the field in a single well) for all temperatures considered; ii)
at all times after the quench when the temperature is low; iii) at very early times after the
quench when the temperature is high. We start by considering the  system before the quench is
implemented ($\alpha=0$), that is, the dynamics with the potential $V(\phi) = \frac{1}{2}\phi^2
+ \frac{1}{8}\phi^4$ \cite{boyanovsky:03}.

A simple way to  implement the HHA in this situation is to factor the cubic term in eq.
\ref{lang} as $\phi^3({\bf x},t)\to3\langle\phi^2\rangle\phi({\bf x},t)$, where
$\langle\phi^2\rangle$  is the translationally invariant mean-square variance of the field.
Since the field is in thermal equilibrium, $\langle\phi^2\rangle$ is time-independent.  In this
approximation, the equation of motion becomes linear, \begin{equation}
\ddot\phi-\nabla^2\phi=-\left(1+\frac{3}{2}\langle \phi^2\rangle\right)\phi. \end{equation} The
Fourier transform of the field is a complex quantity given by \begin{equation}  \bar{\phi}({\bf
k},t)=\frac{1}{(2\pi)^2}\int\!d\,^2x\,\phi({\bf x},t)\,e^{i{\bf k\cdot x}}. \label{ffphi}
\end{equation} The corresponding equation of motion for $\bar{\phi}({\bf k},t)$ in $k$-space is
[with $k=(k_x^2+k_y^2)^{1/2}$] \begin{equation} \ddot{\bar{\phi}}=-\omega^2(k)\bar{\phi},
\label{eomk} \end{equation} with the effective frequency squared \begin{equation}
\omega^2(k)=k^2+1+\frac{3}{2}\langle \phi^2\rangle=k^2+m_H^2. \label{w2} \end{equation} We have
introduced the Hartree mass $m_H^2=1+\frac{3}{2}\langle\phi^2\rangle$. Notice that when there
are fluctuations in this field ($\langle\phi^2\rangle>0$) the effective Hartree mass is  greater
than the ``bare'' mass ($m^2=1$) given by the potential in eq. \ref{V}. In this approximation,
the modes evolve independently of one another, just as they would in a linear truncation of the
equation of motion, only now each interacts with the homogeneous mean-field  background. A 
higher-order approximation is needed, however, to implement direct scattering between different
modes.

If the system is in thermal equilibrium, the momentum modes in $k$-space will satisfy
\begin{equation} \langle |\bar{\pi}({\bf k})|^2\rangle=T. \label{pimode} \end{equation} It then
follows from eq. \ref{eomk} that the corresponding power spectrum for the field is
\begin{equation} \langle |\bar{\phi}({\bf k})|^2\rangle=\frac{T}{k^2+m_H^2}. \label{phimode}
\end{equation} To obtain the equilibrium quantity $\langle\bar\phi^2\rangle$, one integrates eq.
\ref{phimode} over the entire $k$-space and divides by the area. (Equivalently, one can
integrate the corresponding two-point correlation function over $x$-space to obtain 
$\langle\phi^2\rangle$, since Parseval's  theorem guarantees that the two quantities are the
same.) One must be careful here, since the lattice results depend logarithmically on the
ultraviolet cutoff (lattice-spacing). We refer the  reader to ref. \cite{borrill}, where a
method is provided to handle such dependence perturbatively.

\begin{figure} \includegraphics[width=3in,height=3in]{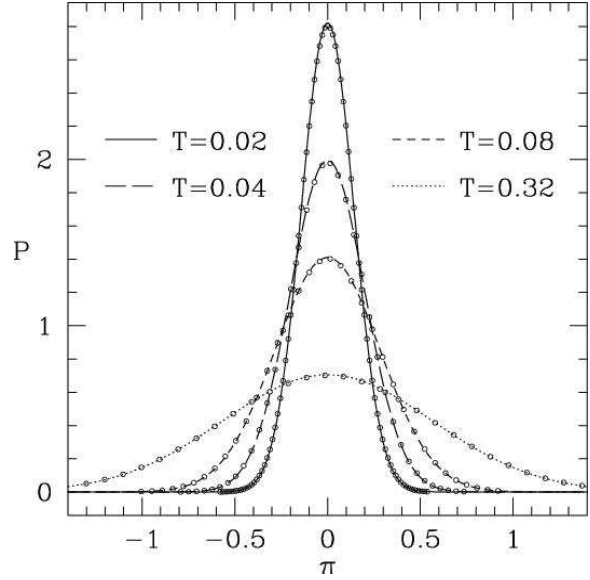} \caption{The probability
distribution of $\pi$ for various temperatures. The continuous lines are  the theoretical
predictions given by eq. \ref{P_pi}.} \label{dist_pi} \end{figure}

\begin{figure} \includegraphics[width=3in,height=3in]{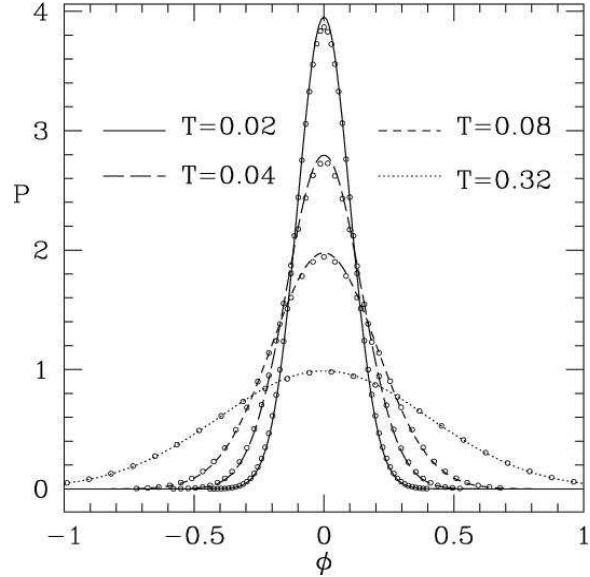} \caption{The probability
distribution of $\phi$ for various temperatures. The continuous lines are  the approximations
given by eq. \ref{P_phi}.} \label{dist_phi} \end{figure}

\begin{figure} \includegraphics[width=3in,height=3in]{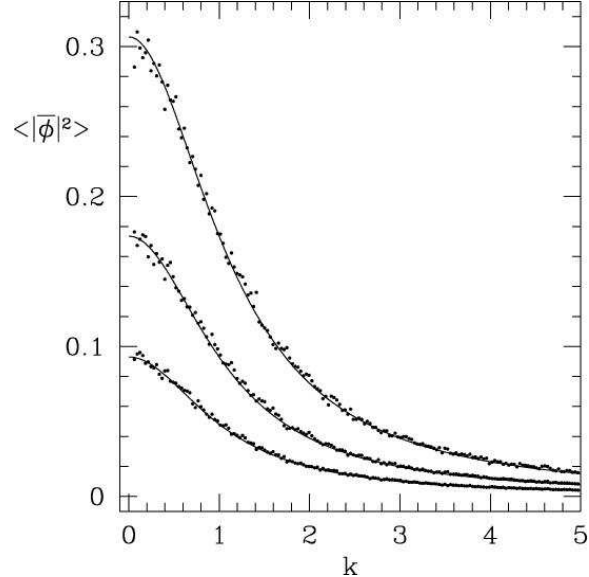} \caption {The power spectrum
$\langle |\bar{\phi}(k)|^2\rangle$ for temperatures $T$=0.1,  0.2, and 0.4. The solid lines are
the approximations given by eq. \ref{phimode}.} \label{spec_phi} \end{figure}

In figures \ref{dist_pi} and \ref{dist_phi} we show the probability distributions for the
initial  state of the field and its momentum  for various temperatures. These distributions are
obtained numerically by collecting the values  of $\pi({\bf x},0)$ and $\phi({\bf x},0)$ of
every microstate into bins of width $\Delta\pi=\Delta \phi=0.01$, and then normalizing so that
the sum over each distribution is unity. Not all data points are shown, making these  figures
more readable. Also shown are the Gaussian  probability distributions (solid lines) that are
predicted from the Boltzmann-Gibbs formalism, \begin{eqnarray} \label{P_pi}
P(\pi)=&\frac{1}{\sqrt{2\pi T}}e^{-\frac{\pi^2}{2T}} \\ P(\phi)=&\frac{1}{\sqrt{2\pi
aT}}e^{-\frac{\phi^2}{2aT}}. \label{P_phi} \end{eqnarray} The fitting parameter $a\simeq 0.51$
is obtained by comparing the area (and ensemble)-averaged two-point corrrelation function of
$\phi$. The variance in figure \ref{dist_phi},  $\langle\phi^2\rangle$, can be fitted linearly,
$\langle\phi^2\rangle = aT$. The agreement between the numerical results and the theoretical
predictions  confirms the accuracy of the lattice dynamics and the applicability of the HHA for
low enough temperatures.  The probability distribution of $\pi({\bf x},0)$ is exact, while  that
of $\phi({\bf x},0)$ is indeed close to Gaussian for the temperatures considered.  We also show,
in figure \ref{spec_phi}, the power spectrum of the field modes,  $\langle |{\bar \phi}({\bf
k},0)|^2\rangle$, and compare it with the prediction from eq. \ref{phimode}, valid for the HHA.

\subsection{Dynamical evolution and breakdown of Hartree approximation} In this subsection we
examine the dynamics of the field after the quench. We will limit our investigation to the case
where $\alpha\rightarrow \alpha_c=3/2$ at $t=0$, that is, the  potential is switched from a
single well to a SDW. The dynamics for ADW potentials will be examined in section 5.

At $t=0$ the coupling to the bath is set to zero, and the dynamics is thus conservative
thereafter. The system is closed, and can be described by a microcanonical ensemble of energy
$\langle E\rangle  \pm \delta E$, where $\delta E/\langle E \rangle \ll 1$. As we remarked
before, for large enough systems in equilibrium, as is the case here for $t < 0$, a description
with fixed $E$ is equivalent to one with fixed $T$. We will use this initial temperature to
characterize the system. It is interesting to note that in the thermodynamic limit, an
instantaneous quench to {\it any} positive value of $\alpha$ will not change the energy of the
system.  The field is initially symmetric in its distribution and the quench introduces an odd
term into the Hamiltonian.  Instead, the energy is redistributed between its kinetic, gradient,
and potential parts, and as a result the final equilibrium temperature will necessarily differ
from the initial temperature. In this investigation, however,  this change is always less than
$0.15$ percent. For open systems, where the system remains in contact with a heat bath, we refer
the reader to ref. \cite{gleiser-howell}.

Due to the initial thermal distribution, defined entirely by the temperature $T$, the sudden 
introduction of the cubic term into the potential actually has a minor effect  on most of the
field. In general, there is little  net change in the forces present throughout the system,
since the curvature of the single- and double-well potentials are the same at $\phi=0$ and the
majority of the field is indeed close to this value (see figure 2).  That is, the linear
equation of motion remains unchanged for the  small-amplitude fluctuations that represent the
peak of the field's probability distribution. The  few locations where the field is more
responsive to the perturbation correspond to regions with relatively large absolute values of
$\phi$. These values define the  tails of  the probability distribution. The new nonlinear term
present in the potential breaks the symmetry in the equation of motion, and this second-order
effect causes larger-amplitude field  fluctuations with negative (positive) values to experience
a larger (smaller)  restoring force than before. Overall, a net positive force is imposed on the
symmetrically prepared field and, as a consequence, its average starts at $\langle\phi\rangle=0$
and then begins to move toward $\phi_{\rm max}$, the maximum of the potential barrier between
the two states.

We will see that $\langle\phi(t)\rangle$ is  initially driven more out of equilibrium than all
of the fluctuations about it. This is not surprising, as longer wavelength modes have longer
relaxation time-scales \cite{langer}. It is  natural, then, to distinguish the dynamics of the
two from each other. First, define $\phi_{\rm ave}(t)\equiv\langle\phi(t)\rangle$ and let
$\delta\phi({\bf x},t)$ represent the fluctuations about this average. Then, \begin{equation}
\phi({\bf x},t)=\phi_{\rm ave}(t)+\delta\phi({\bf x},t), \label{sepphi} \end{equation} where
$\langle\delta\phi({\bf x},t)\rangle=0$ by definition. In the HHA, one assumes that these
fluctuations and their momenta remain Gaussian in their distribution.  One then obtains an 
equation of motion for $\phi_{\rm ave}(t)$, \begin{equation} \ddot\phi_{\rm ave}=-\frac{\partial
V_{\rm eff}}{\partial\phi_{\rm ave}}~, \label{eomphiave} \end{equation} where we have introduced
the effective potential for $\phi_{\rm ave}$, \begin{equation} \label{Veff} V_{\rm
eff}\left(\phi_{\rm ave},\langle\delta\phi^2\rangle\right)=  \left[1-m_{\rm
H}^2(t)\right]\phi_{\rm ave} +\frac{1}{2}\,m_{\rm H}^2(t)\,\phi_{\rm ave}^2  -\frac{\alpha_{\rm
c}}{3}\,\phi_{\rm ave}^3+\frac{1}{8}\,\phi_{\rm ave}^4, \end{equation} that depends explicitly
on the mean-square variance of the fluctuations through the time-dependent Hartree mass  $m_{\rm
H}^2(t)=1+\frac{3}{2}\langle\delta\phi^2(t)\rangle$.

\begin{figure} \includegraphics[width=3in,height=3in]{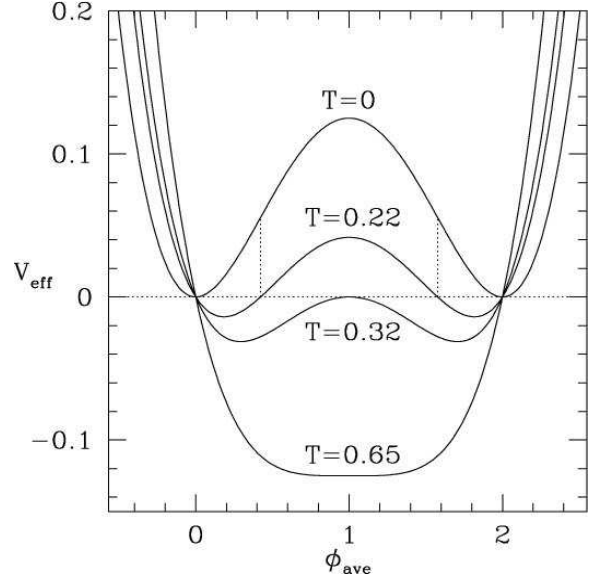} \caption {The effective
potential for $\phi_{\rm ave}$, just after the quench, for various initial temperatures $T$. The
two vertical lines correspond to the inflection points of the bare ($T=0$) potential. They are
also the turning points for the potential at $T\simeq 0.22$. At $T\simeq 0.32$ the turning point
is $\phi_{\rm max}$.} \label{V_eff} \end{figure}

$V_{\rm eff}$ at $t=0$ is shown in figure \ref{V_eff} for various temperatures.  For
sufficiently low initial temperatures (how low will become clear soon), $\phi_{\rm ave}(t)$ will
oscillate about the LHS minimum of the effective potential, $\phi_-
=1-\sqrt{1-3\langle\delta\phi^2\rangle}$.   It will then begin to relinquish its energy to
higher modes through nonlinear scattering, and its amplitude will decay. It is possible, then,
to approximately bound this region of oscillation from the set of turning points for  $\phi_{\rm
ave}(t)$ [with $\langle\delta\phi^2\rangle$ retaining its $t=0$ value], \begin{equation}
\phi^{\rm tp}_{\rm ave}=\left\{0,\,1\pm\sqrt{1-6\langle\delta\phi^2\rangle},\,2\right\}.
\label{tp} \end{equation} These values give the location of $\phi_{\rm ave}(t)$ where $V_{\rm
eff}=0$, and  $\pi_{\rm ave}(t)=\partial_t\phi_{\rm ave}(t)$ vanishes. When
$\langle\delta\phi^2\rangle<1/6$ there are four such turning points because of the symmetry
about $\phi_{\rm max}$. The primary region of  interest is when $\phi^{\rm tp}_{\rm
ave}\simeq\phi_{\rm inf(-)}$, when fluctuations begin to probe the spinodal region of $V_{\rm
eff}$. By using eq. \ref{tp}, we can predict that $\phi_{\rm ave}(t)$ will probe this region
when $\langle\delta\phi^2\rangle=1/9$, or $T\simeq 0.22$. Clearly, this is also where the
leading-order Hartree approximation fails.

It is interesting to consider also the case when $\langle\delta\phi^2\rangle=1/3$, or
$T\simeq0.65$. There are only two turning points for $\phi_{\rm ave}$, since $V_{\rm eff}$ is
now a single-well potential symmetric about $\phi=\phi_{\rm max}$. The fluctuations are  large
enough that the local maximum in the bare potential has little influence in the dynamics, and
$\phi_{\rm ave}=\phi_{\rm max}$. The fluctuations actually probe beyond both  minima of the
potential. This case is similar to the phenomena of symmetry restoration in continuous  phase
transitions, and so in this context $T\simeq0.65$ would be considered the critical temperature,
$T_c$.  For all higher temperatures, $\phi_{\rm ave}=\phi_{\rm max}$ \cite{phase,gleiser-mix}.
In this work, we are only concerned with temperatures well below $T_c$, in fact below $T=0.32$,
where $\phi^{\rm tp}_{\rm ave} = \phi_{\rm max}$ (see figure 4).

In order to further quantify the dynamics in the HHA, we derive the set of differential
equations for the time evolution of the coupled two-point correlation functions.  We define the
translationally invariant quantities \begin{eqnarray} \nonumber G_{\phi\phi}({\bf x-y},t) &=&
\langle\delta\phi({\bf x},t)\delta\phi({\bf y},t)\rangle \\ G_{\pi\pi}({\bf x-y},t) &=&
\langle\delta\pi({\bf x},t)\delta\pi({\bf y},t)\rangle \\ \nonumber G_{\pi\phi}({\bf x-y},t) &=&
\frac{1}{2}\left(\langle\delta\pi({\bf x},t) \delta\phi({\bf y},t)\rangle+\langle\delta\phi({\bf
x},t) \delta\pi({\bf y},t)\rangle\right). \label{Gcorr} \end{eqnarray} Upon taking the
time-derivative of each expression, the equations of motion can then  equivalently be written in
$k$-space. They are \begin{eqnarray} \nonumber \dot G_{\phi\phi}(k,t) &=& 2G_{\pi\phi}(k,t) \\
\dot G_{\pi\pi}(k,t) &=& -2\bar\omega^2 G_{\pi\phi}(k,t) \\ \nonumber \dot G_{\pi\phi}(k,t) &=&
G_{\pi\pi}(k,t)-\bar\omega^2 G_{\phi\phi}(k,t)~, \end{eqnarray} where ${\bar \omega}^2(k) = k^2
+ V_{\rm eff}'' \left [\phi_{\rm ave}(t),\langle\delta{\bar \phi}^2(t)\rangle \right ]$.

These equations, coupled with \begin{eqnarray} \dot\pi_{\rm ave} &=& -\frac{\partial V_{\rm
eff}}{\partial\phi_{\rm ave}} \\ \nonumber\dot\phi_{\rm ave} &=& \pi_{\rm ave}, \end{eqnarray}
give the full dynamics of $\pi_{\rm ave}(t)$ and $\phi_{\rm ave}(t)$ and their corresponding
two-point correlation functions  $\langle\delta\pi^2({\bf k},t)\rangle$ and
$\langle\delta\phi^2({\bf k},t)\rangle$ in the HHA. These two groups of expressions are likewise
coupled through the quantities
$\langle\bar{\delta\phi}^2(t)\rangle=\langle\delta\phi^2(t)\rangle$ and $\phi_{\rm ave}(t)$. The
initial conditions  are given by eqs. \ref{pimode}, \ref{phimode}, and also $\pi_{\rm
ave}(0)=\phi_{\rm ave}(0)=0$. 

These five first-order equations were solved numerically using a fourth-order Runge-Kutta
algorithm. The same lattice spacing from the leap-frog algorithm was used. The calculation of
$\langle\bar{\delta\phi}^2(t)\rangle$ is obtained by numerically averaging the radial function
$G_{\phi\phi}(k,t)$ over the disk in $k$-space with radius $k^2=k_x^2 + k_y^2$, bounded by 
$k_{\rm max}=\pi/\Delta x$. There is a slight discrepancy between the area of rectangular phase 
space used in the two-dimensional numerical integrations and the circular phase space used  in
this method. In light of the previous discussion regarding the dependence of the correlation
functions on the lattice spacing, this difference could possibly introduce error in the 
calculations of $\langle\bar{\delta\phi}^2(t)\rangle$. To compensate for the smaller area used
in the HHA, the initial conditions for $\langle\bar{\delta\phi}^2(k,0)\rangle$, given by eq.
\ref{phimode}, are multiplied by a normalization factor of 1.041 so that the radial integration
gives the expected average $\langle\bar{\delta\phi}^2\rangle=aT$ at $t=0$. The results will
confirm the validity of this  normalization method.

\begin{figure} \includegraphics[width=3in,height=3in]{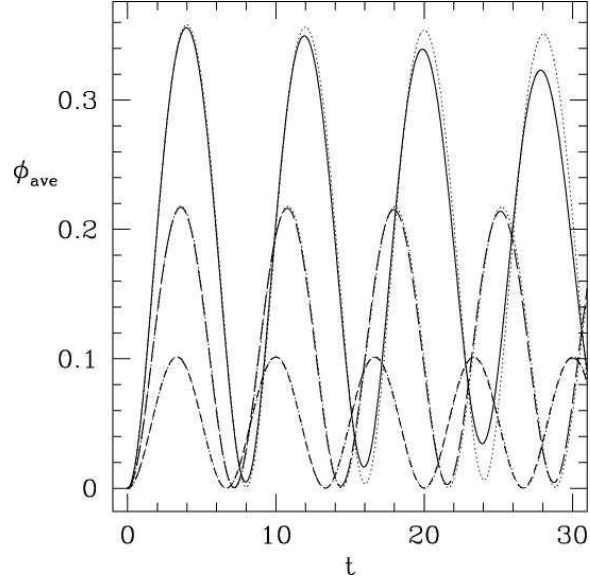} \caption[A comparison
of numerical and Hartree dynamics for $\phi_{\rm ave}(t)$] {Numerical evolution of $\phi_{\rm
ave}(t)$ compared to that obtained through the Hartree approximation, for various initial
temperatures. Shown are $T=0.06$ (short dash), $T=0.12$ (long dash), and $T=0.18$ (solid), as
well as the corresponding Hartree  approximations (dotted).} \label{phiavehartree} \end{figure}

Shown in figure \ref{phiavehartree} is the numerical time evolution of $\phi_{\rm ave}(t)$ 
compared to that obtained through the HHA, for various initial temperatures. The agreement is
clearly excellent at lower temperatures, where $\langle\phi_{\rm ave}(t)\rangle$  undergoes
almost-linear oscillations about the minimum of $V_{\rm eff}$: there is very little scattering
between modes. At $T\simeq 0.18$ scattering begins to take place, allowing for a slow transfer
of energy from low-$k$ modes to higher-$k$ modes not accounted for in the HHA.  We will soon
show that this is also the temperature where the synchronous emergence of spatio-temporal order
begins to take place. Notice also the dependence of the oscillation frequencies on the initial
thermalization temperature. 

\begin{figure} \includegraphics[width=3in,height=3in]{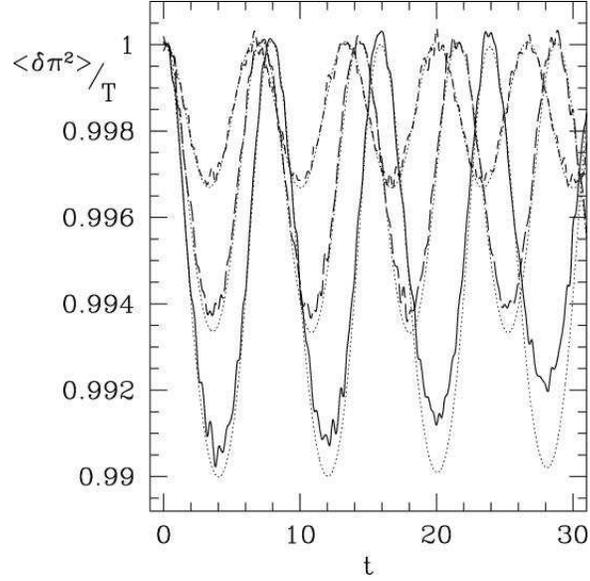} \caption {The
numerical evolution of $\langle\pi^2(t)\rangle/T$ compared to that obtained  through the Hartree
approximation, for various initial temperatures. Shown are $T=0.06$ (short dash), $T=0.12$ (long
dash), and $T=0.18$ (solid), as well as the corresponding Hartree  approximations (dotted).}
\label{corrpihartree} \end{figure}

\begin{figure} \includegraphics[width=3in,height=3in]{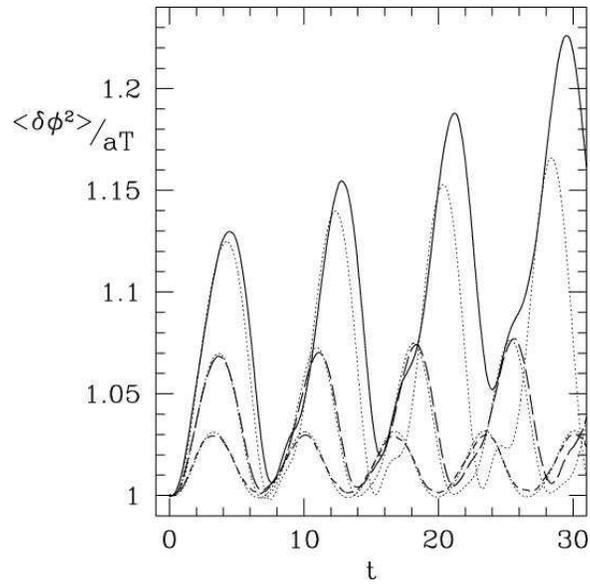} \caption {The
numerical evolution of $\langle\phi^2(t)\rangle/aT$ compared to that obtained  through the
Hartree approximation, for various initial temperatures. Shown are $T=0.06$ (short dash),
$T=0.12$ (long dash), and $T=0.18$ (solid), as well as the corresponding Hartree  approximations
(dotted).} \label{corrphihartree} \end{figure}

Figures \ref{corrpihartree} and \ref{corrphihartree} compare the numerical  time-evolution of
$\langle\delta\pi^2(t)\rangle$ and $\langle\delta\phi^2(t)\rangle$ with the corresponding HHA
predictions. The quantities are normalized to unity at $t=0$ through $\langle\pi^2\rangle=T$
and  $\langle\phi^2\rangle = aT$, respectively. The agreement is again excellent at lower
temperatures. At $T=0.18$, however, it is evident  that energy is being fed into the
fluctuations of the field. The full numerical evolution depicts the amplitude of oscillation of
$\langle\delta\pi^2(t)\rangle$  decreasing at a rate larger than that predicted by the HHA.
Similarly, the amplitude of oscillation of  $\langle\delta\phi^2(t)\rangle$ increases at a
larger rate. Again, this discrepancy is due  primarily from the approximation's inability to
account for direct scattering between modes. 

\begin{figure}[t] \centering \includegraphics[width=4in,height=4in]{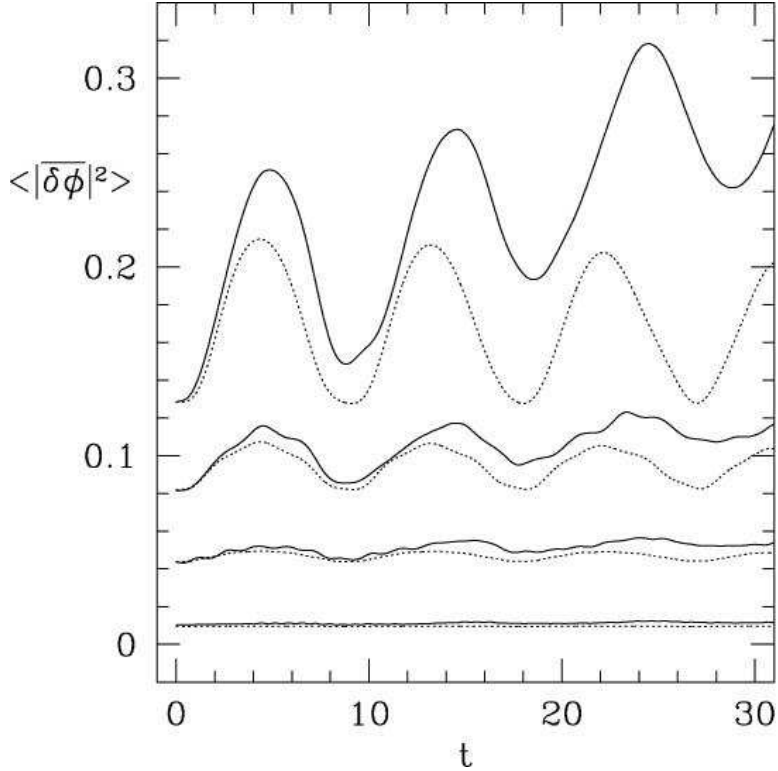}
\renewcommand{\baselinestretch}{1.0} \caption[The two-point correlation function for various
values of $k$] {The numerical evolution of the two-point correlation function 
$\langle|\bar{\phi}(k,t)|^2\rangle$ for various values of $k$.  Also shown are the results from
the corresponding Hartree approximation (dotted  lines). The initial temperature is $T=0.22$.
The amplitudes decrease for increasing $k$. Shown, from the top down, are $k$=0.75, 1.25, 2.0,
and 4.75.} \label{speckhartree} \end{figure}

In figure \ref{speckhartree} we show the transfer of energy between different  $k$-modes.
Displayed is the time evolution of $\langle|\bar{\delta\phi}(k,t)|^2\rangle$ for indicative
values of $k$. The initial thermalization temperature is $T=0.22$, corresponding to the value in
which the majority of the field probes the spinodal region (cf. section IV). This is well above
the region of temperatures in which the HHA is accurate. From the figure it is clear that
scattering between low $k$ modes modes is taking place.  The approximation is only accurate for
shorter wavelength modes, due to the fact that they remain in equilibrium even after the quench.
The breakdown of the HHA at longer wavelengths implies that important nonlinear effects are
strongly influencing the behavior of the field. In the next section we investigate the emerging
dynamics of these modes.

\section{Emergence of spatio-temporal order in a SDW model}

\subsection{Beyond the Homogeneous Hartree Approximation}

The results of the last section should have made it clear that at high temperatures nonlinear
scattering leads to an efficient energy transfer between modes. This energy transfer will
eventually lead to the thermalization of the field in the SDW potential. As predicted by the
HHA, at $T\gtrsim 0.18$ there is consistent energy transfer from the zero-mode to these
higher-$k$ modes. At these initial temperatures,  $\phi_{\rm ave}(t)$ undergoes damped
oscillations about its effective minimum. For $T=0.22$, the amplitudes of these oscillations are
large enough that $\phi_{\rm ave}(t)$ just crosses the inflection point of the bare potential:
roughly half of the  field probes the spinodal region.

\begin{figure} \includegraphics[width=3in,height=3in]{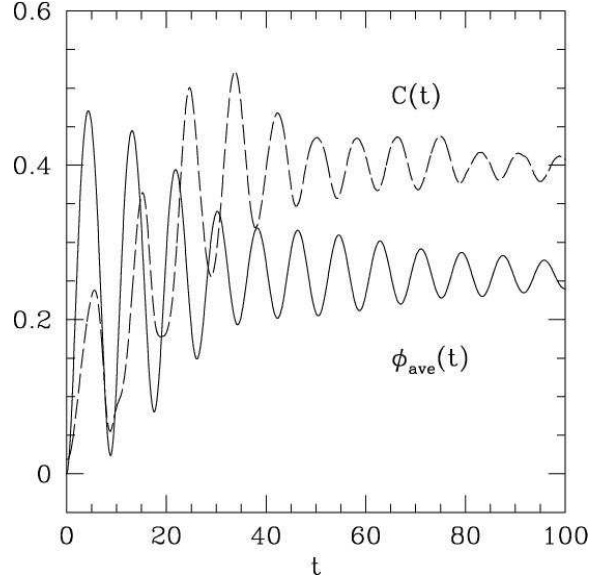} \caption {$\phi_{\rm
ave}(t)$ (solid line) and $C(t)$ (dashed line) for $T=0.22$.} \label{corrandave} \end{figure}

Figure \ref{corrandave} illustrates the transfer of energy from $\phi_{\rm ave}(t)$ to the
fluctuations $\langle\delta\phi^2(t)\rangle$ for  $T=0.22$. The quantity \begin{equation}
C(t)\equiv \frac{\langle\delta\phi^2(t)\rangle-aT}{aT} \end{equation} is introduced so that a
visual comparison can be made between these two components of the field.  Notice the correlation
between the decaying amplitude of oscillation for $\phi_{\rm ave}(t)$ and the increasing
amplitude of the fluctuations. At $t\simeq 60$ most of this energy transfer is complete, after
which time the two quantities display smaller oscillations that are almost exactly out of phase
with each other. For earlier times, $0\leq t\leq 60$, a highly nonlinear scattering process
occurs, as the zero-mode of the field pumps its energy into its neighboring small-$k$ modes.

\begin{figure} \includegraphics[width=3in,height=3in]{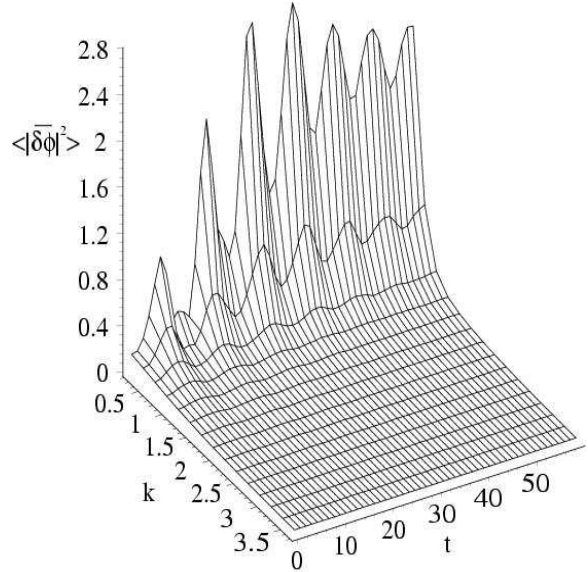} \caption {The spectrum of the
field for times $0\leq t\leq 60$.} \label{spec60} \end{figure}

Figure \ref{spec60} shows the numerical time-evolution of
$\langle|\bar{\delta\phi}(k,t)|^2\rangle$  for $0\leq k\leq 4$ and $0\leq t\leq 60$, at
$T=0.22$. First, notice that growth occurs mainly for modes with $0\leq k\leq 0.5$. Their
maximum amplitude reaches a value that is up to 15 times greater than at $t=0$. The oscillations
then begin to decay and, for $t>60$ (discussed later), the amplitudes of the modes take on a
constant value, albeit still many times larger than their initial value. It is also apparent
that for the time interval shown  modes with $k>1$ are nearly time-independent, maintaining
their initial thermal spectrum given by eq. \ref{phimode}. This is true of all large-$k$ modes
present in the numerical simulation, up to $|k|_{\rm max}\simeq 15.7$, which also indicates that
the lattice spacing used has no effect on the emerging dynamics of the system.

We would now like to search for possible spatio-temporal ordering during the field's approach to
equilibrium. However, due to the noisy initial conditions, it is quite difficult to extract
information about any emergent pattern, be it spatial and/or temporal.  An added advantage to
having an analytical expression for the initial thermal spectrum for all
temperatures of interest (cf. eq. \ref{phimode}) should be clear: a Wiener filter is extremely 
successful at
removing undesirable noise whenever its spectrum is known, all the while maximally preserving the
nonperturbative fluctuations that emerge \cite{numericalrecipes}.

\subsection{Synchronous emergence of spatio-temporal order}

With the filtered field, one can catalog and track the location of all local extrema in the
system at each instant in time and for all initial temperatures. To distinguish the 
large-amplitude fluctuations from the long-wavelength thermal noise that  is also passed by the
filter, only the values of these extrema that are greater than the characteristic amplitude of
the noise, $\delta\phi^2({\bf x},t)>aT$, are considered.  With ample sorting, a library is
compiled containing the description of every large-amplitude fluctuation present during the
entire evolution of the system (the maximum time for the simulations is $t_{\rm max}=2000$). For
each fluctuation the following attributes are recorded:  its core location ${\bf x}_{\rm f}$,
nucleation time $t_{\rm nuc}$, core amplitude $\phi_{\rm a}$ (measured at ${\bf x}_{\rm f}$),
radius $R$  (measured at half maximum), its total number of oscillations $N_{\rm osc}$, the
period of each  oscillation ${\cal T}_N$, and its lifetime $\tau$. 

Each of these attributes is time-dependent, with the exception of $N_{\rm osc}$ and $\tau$.
Almost  every large-amplitude fluctuation oscillates anharmonically about the background  value
$\phi_{\rm ave}(t)$, which is recognized by the fact that most of them reappear at approximately
the same location with regular frequency. Therefore, in general it is possible to  establish the
relationship $N_{\rm osc}=\tau/{\cal T}$, where  ${\cal T}\equiv\langle{\cal T}_N\rangle$ is
the  average period of oscillation. The set of observables is further reduced by averaging ${\bf
x}_{f}$, $\phi_{\rm a}$, and $R$ over their corresponding values at the maximum of each
oscillation for the lifetime of the fluctuation. Other than the nucleation time $t_{\rm nuc}$,
each attribute is now time-independent. Hereafter, we will omit the $\langle\cdots\rangle$
notation and assume that these values are indeed averages.

With this information, a probability distribution function for the large-amplitude  fluctuations
is obtained for each initial temperature. This function, denoted as \linebreak $F_T({\bf x}_f,
t_{\rm nuc}, \phi_{\rm a}, R, {\cal T}, \tau)$, can be used to calculate averages over any of
its variables. Similarly, one can integrate this function over a subset of variables to obtain a
reduced distribution. In all cases, $F_T$ is normalized  so that a numerical integration over
all variables evaluates to unity. The bin widths of each  variable (as any one variable of $F_T$
will be visualized as a histogram) will be specified below.

\begin{figure} \includegraphics[width=3in,height=3in]{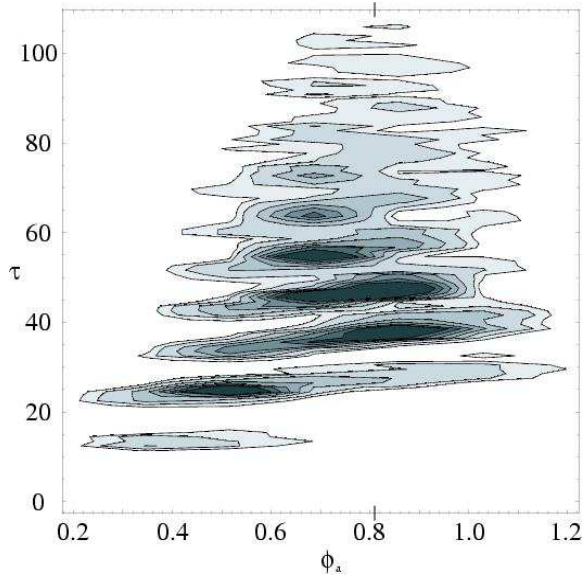} \caption {The 
probability distribution for the amplitude and lifetime of all fluctuations accounted for early 
in the evolution ($t<100$). The darkest regions correspond to the most abundant fluctuations
(probability $f=0.07$), with each lighter region then descending in intervals $\Delta f=0.01$.}
\label{disttauphi} \end{figure}

For the moment, we focus our attention on the characteristics of fluctuations early in the
evolution of the field, $0\leq t\leq 100$, and at an initial temperature $T=0.22$.  The quantity
\begin{equation} f_{T}(\phi_{\rm a},\tau)=\int_{A}\!\!d\,^2 x_{\rm f}\int_{0}^{100}\!\!\!dt_{\rm
nuc} \int_{0}^{L/2}\!\!\!dR\int_{0}^{t_{\rm max}}\!\!\!d{\cal T}\,\,F_{T} \end{equation} gives
the reduced distribution function for the core amplitude and lifetime, and is  shown in figure
\ref{disttauphi}. The bin widths are $\Delta\phi_{\rm a}=0.05$ and $\Delta\tau=1$. The darkest
regions in the figure correspond to the most probable fluctuations. Particularly noticeable is
that fluctuations with core amplitudes $0.55\leq \phi_{\rm a}\leq1.0$ and lifetimes
$35\leq\tau\leq65$ dominate the distribution. One can also see that fluctuations with small
amplitudes $0.2\leq\phi_{\rm a}\leq 0.5$ have mostly short lifetimes. There is a  group of
fluctuations, however, that have lifetimes $80\leq\tau\leq100$ and thus remain present in the
system for substantially long periods of time. In addition, their core amplitudes are confined
to a narrower region $0.7\lesssim\phi_{\rm a}\lesssim 0.9$. Also marked on the horizontal axes
is the location $\phi_{\rm a}=0.81$, which corresponds to the difference between the local
maximum and the left minimum of the effective potential for $T=0.22$, given by eq. \ref{Veff}.
It follows that these long-lived oscillatory fluctuations have large enough amplitudes that
their cores probe a  narrow region centered about $\phi_{\rm max}$. In fact, the large majority
of long-lived fluctuations probe beyond the inflection point of $V_{\rm eff}$ at $\phi_{\rm
inf(-)}\simeq 0.4$.  They are clearly nonperturbative in nature and would not be seen within the
HHA.

\begin{figure} \includegraphics[width=4in,height=8in]{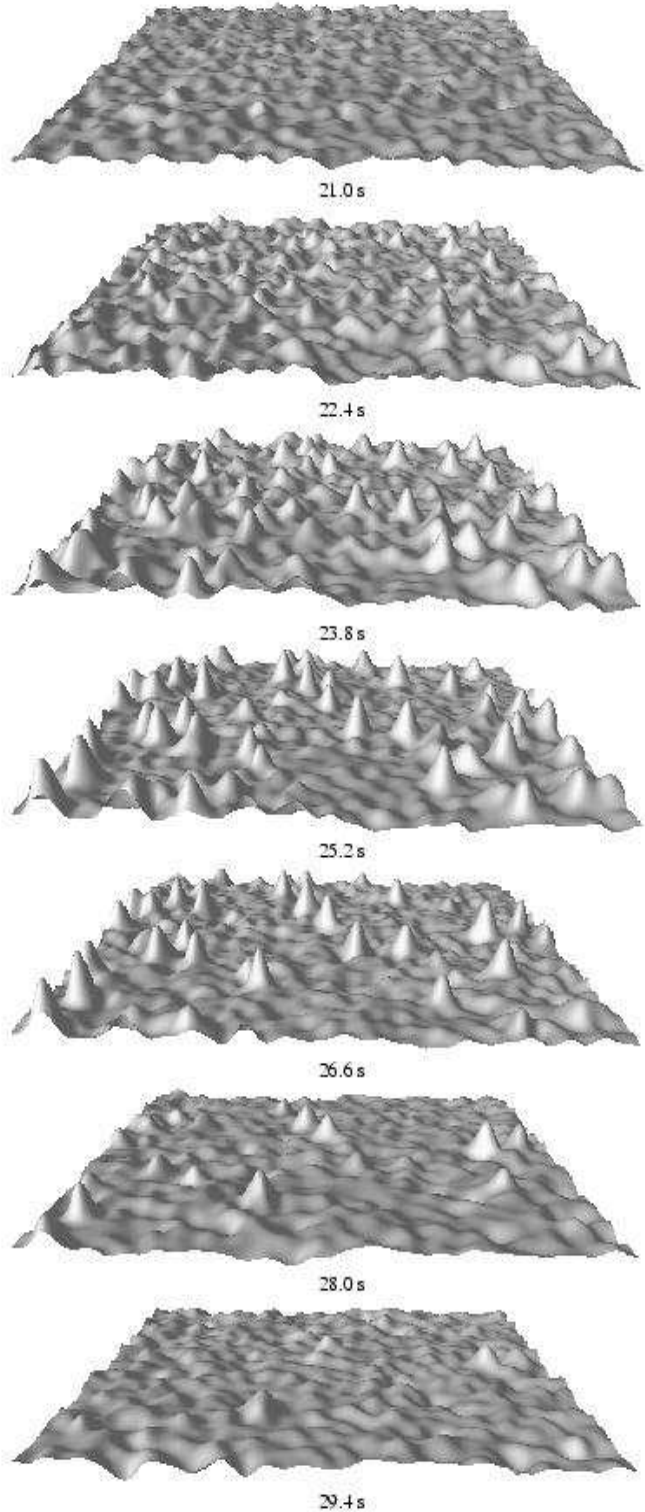} \caption {Snapshots
displaying the synchronous emergence of oscillons in the two-dimensional field. The initial
temperature is $T=0.22$. Simulations can be viewed in ref. \cite{website}.} \label{oscsnap}
\end{figure}

In figure \ref{oscsnap} we show a sequence of snapshots of the filtered field within the time
interval $21.0 \leq t \leq 29.4$. There are several features to be noted. [Clearly, they are
more striking with a full animation. We invite the reader to view simulations in ref.
\cite{website}.] First, large-amplitude fluctuations seem to be of similar size (spatially-bound
configurations)  and spread throughout the lattice. Second, they emerge in near synchrony.  As
we will see below, this global ordering persists for the first few oscillation cycles. Third, as
can be seen from figure \ref{disttauphi}, these configurations can be quite long-lived.

We identify these spatially-bound, long-lived configurations with oscillons, which have been
originally found as deterministic solutions of models with amplitude-dependent nonlinearities
\cite{gleiser,bettinson}. In these studies, the field was  prepared with a bubble-like profile
(typically a Gaussian or a tanh), and for large enough initial radius and amplitude, observed to
fall into an oscillon configuration, characterized by a rapid oscillation of its core, as shown
in figure \ref{oscsnap}. Using harmonic perturbation analysis about an oscillon solution,  it
has been shown that such configurations require a minimum radius (in 2 spatial dimensions for
the SDW potential used here) \cite{sornborger}, \begin{equation} R_{\rm
crit}=\frac{\sqrt{6}}{m}~. \label{Rc} \end{equation}  We will now extend this result to the
general ADW potential. First, we parameterize the radially-symmetric oscillon configuration by a
Gaussian \cite{gleiser, sornborger} \begin{equation} \label{osc_ans} \phi_{\rm osc}(r,t) =
\phi_a(t)\exp[-r^2/R^2],~ \end{equation} where $\phi_a(t)$ gives the time behavior at the core
of the configuration and $R$ is its radial dispertion, assumed here to be constant. (This
assumption is equivalent to reducing the dynamics to one degree of freedom; it has been shown to
provide results in excellent agreement with numerical simulations \cite{gleiser,sornborger}.)
With this ansatz, a Lagrangian can be obtained and, from it, the equation of motion for
$\phi_a(t)$. Linear perturbations about the solution, $\delta \phi = \phi_a(t) - \phi_0(t)$,
oscillate with frequency \begin{equation} \label{fluct_freq} \omega^2(\phi_0,\alpha,R) =
\frac{2}{R^2} + 1 - \frac{4}{3}\alpha\phi_0 + \frac{3}{4}\phi_0^2~. \end{equation} Note that
$\omega^2$ has a minimum at $\phi_0^c=\frac{8}{9}\alpha$. There is an unstable bifurcation  at 
\begin{equation} \label{unst_rad} R_{\rm crit}^2 = \frac{2}{\frac{16}{27}\alpha^2 - 1}~.
\end{equation} For the SDW, $\alpha_c=3/2$ and the instability is given by eq. \ref{Rc}.
Oscillons are only possible for values of $R$ larger than $R_{\rm crit}$. Larger asymmetries
imply smaller values of $R_{\rm crit}$. 

\begin{figure} \includegraphics[width=3in,height=3in]{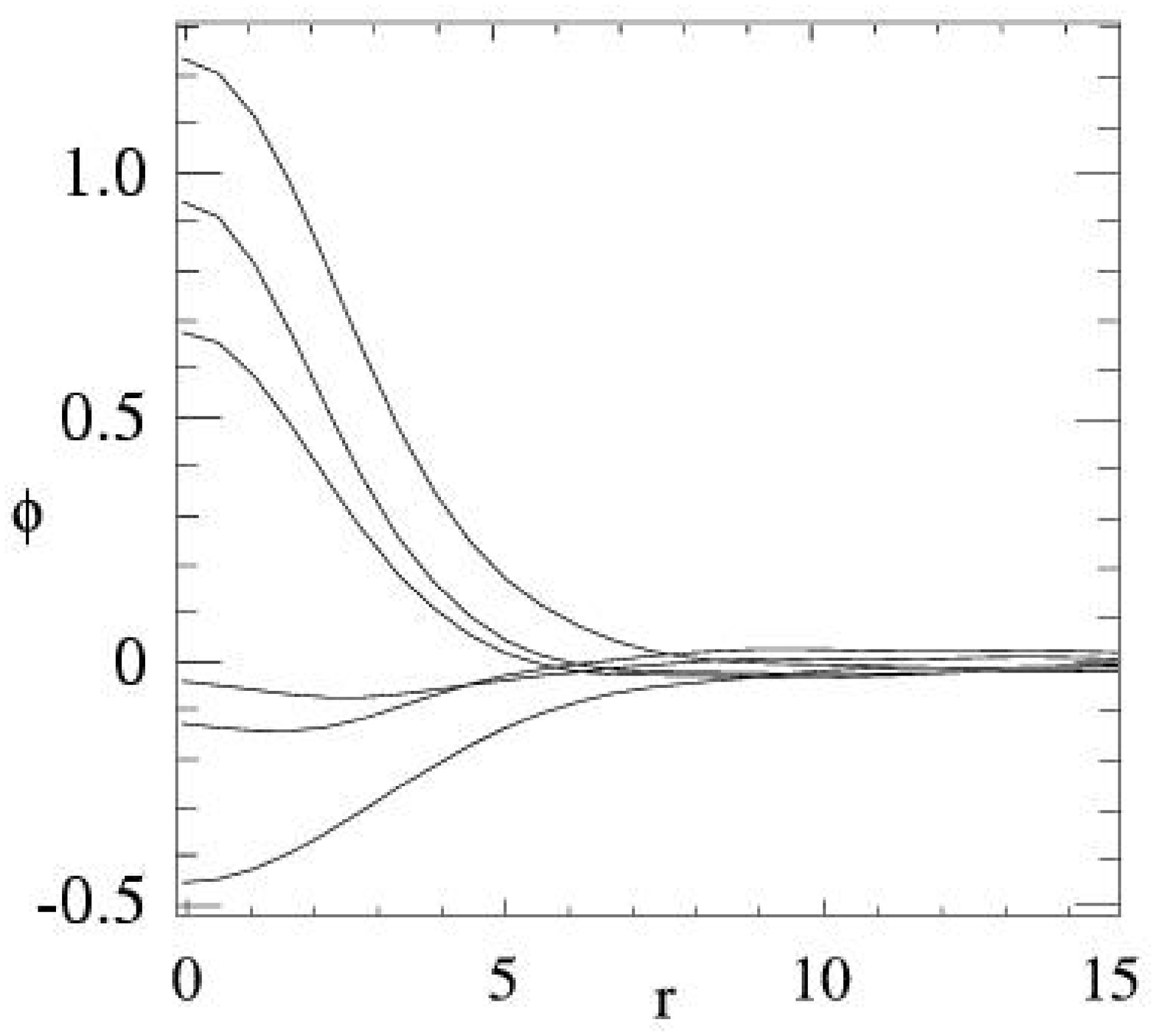} \caption {The profile of a
localized oscillon at sequential time intervals, spanning half its oscillation  period.}
\label{oscprof} \end{figure}

\begin{figure}[t] \centering \includegraphics[width=4in,height=4in]{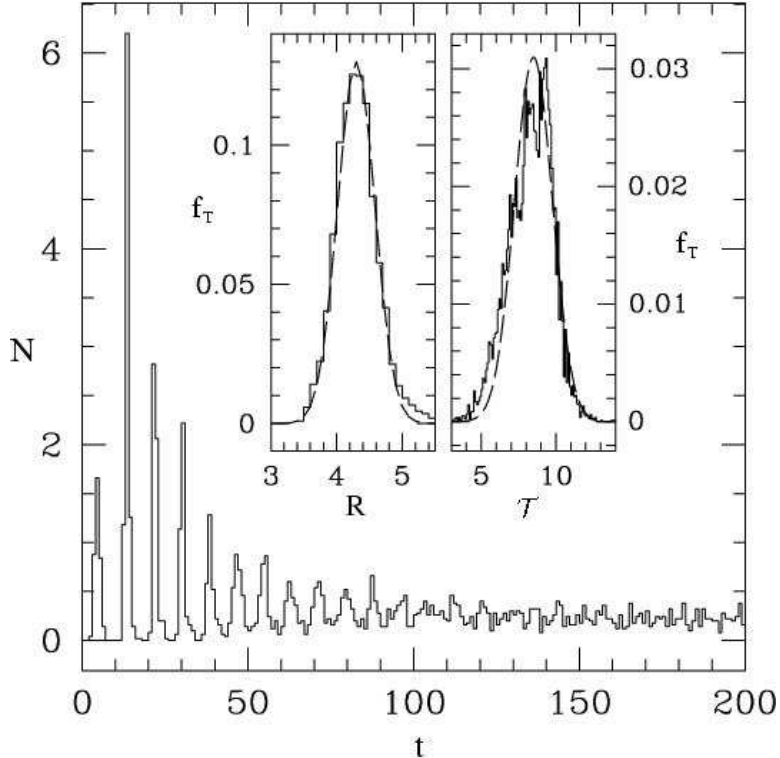}
\renewcommand{\baselinestretch}{1.0} \caption[Various probability distributions for the emergent
oscillons] {The number of oscillons nucleated as a function of time for $T=0.22$. Also shown
(inset) are the probability distributions (solid lines) for the radius $R$ and period ${\cal T}$
of  these oscillons. The fitted curves (dashed lines) are Gaussian functions.} \label{dist}
\end{figure}

In order to strengthen our claim that oscillons are nucleated during the nonequilibrium
evolution of $\phi({\bf x},t)$, the inset in figure \ref{dist} shows the distribution functions 
$f_T(R)$ and $f_T({\cal T})$ for the radii and period of oscillation, respectively. To make a
comparison with deterministic oscillons, only fluctuations with $\phi_{\rm a}\geq 0.81$ (i.e.
those that probe beyond the local maximum of the potential) and $\tau\geq 80$ are considered.
That is, these are the lower bounds placed on the corresponding integrations over $F_T$. The bin
widths are $\Delta R=\Delta {\cal T}=0.1$. The fitted curves are  Gaussian functions with
centers $R_{\rm ave}=4.3$  and ${\cal T}_{\rm ave}=8.5$ and widths $\sigma_R=0.4$ and
$\sigma_{\cal T}=1.8$, respectively. To complete the comparison, consider a deterministic
oscillon evolving in a time-independent effective potential that is representative of this case.
From figure \ref{corrandave} a good approximation would be to take $\langle
C(t)\rangle\simeq0.4$, so that $\langle\delta\phi\rangle\simeq0.15$. The effective mass can
thus  be obtained from the curvature of the left-hand well, given by eq. \ref{Veff}, and
evaluates to $m_{\rm eff}=0.73$. Finally, from eq. \ref{Rc}, we find that the effective minimum
radius for which the deterministic oscillon remains  stable is $R_{\rm crit}^{\rm eff}=
\sqrt{6}/m_{\rm eff}\simeq 3.37$. Comparing this result with figure \ref{dist} we see that,
indeed, the emerging oscillons satisfy this requirement remarkably well. [Note that a proper
comparison between oscillons that evolve in different potentials (i.e. for fields at different
initial temperatures) always requires a  suitable scaling of $\phi$, ${\bf x}$, and $t$.]

Also shown in figure \ref{dist} is the distribution of nucleation times for the emerging
oscillons. This quantity gives the number of oscillons nucleated between $t$ and $t+\Delta t$,
with  $\Delta t=1$. It is obtained by calculating the reduced distribution function $f_T(t_{\rm
nuc})$ integrated over all the remaining variables. The same lower limits are used for the
integrals over $\phi_{\rm a}$ and $\tau$, so that only long-lived large-amplitude fluctuations
are considered. If there are $N_{\rm tot}$ such fluctuations, then the final expression for the
number of oscillons nucleated as a function of time is \begin{equation} N(t)=N_{\rm
tot}f_{T}(t_{\rm nuc}). \end{equation}

The sharp peaks in $N$ at early times, $t<60$, correspond to the synchronous emergence of 
oscillons. For $t>60$, this global ordering gives way to only local ordering, in which oscillons
emerge at  arbitrary times with similar probability. Even this local emergence disappears after
approximately  $t\geq 500$ (more on this later). Notice the correlation between the synchronous
nucleation of oscillons  and the energy loss from the zero-mode (c.f. figure \ref{corrandave})
at early times, $t< 60$.  This phenomenon  was observed for temperatures within $0.16\leq T \leq
0.26$ (see below). Below this temperature range the HHA is valid and no coherent structure
emerges. Above  this temperature range the field separates into large, slowly evolving
thin-walled domains, signalling the approach to criticality. 

\subsection{Synchronous emergence of order is due to parametric resonance}

In order to understand the mechanism behind the observed synchronous emergence, we decompose the
field as in eq. \ref{sepphi}. The linearized equation satisfied by the Fourier modes of the
fluctuations is  \begin{equation} \ddot{\bar{\delta\phi}}=-\bar\omega^2(k)\bar{\delta\phi},
\end{equation} where \begin{equation} \bar\omega(k)^2=k^2+V_{\rm eff}''\left[\phi_{\rm
ave}(t)\right]. \label{omegaparares} \end{equation} In this approximation, the fluctuations obey
a linear equation of motion with a time-dependent frequency determined by the value of
$\phi_{\rm ave}(t)$. Equations of this type, generalized Mathieu equations, are known to exhibit
parametric resonance,  which can lead to exponential amplification ($\sim \exp{\eta t}$) in the
oscillations of $\delta\phi(k,t)$ at certain wavelengths. They have been the focus of much
recent attention in studies of reheating of the universe after an inflationary expansion phase
\cite{kofman}. To verify that this is the mechanism behind the synchronous amplification of
oscillon modes early  in the field's evolution, we use in eq. \ref{omegaparares} the anharmonic
solution for  $\phi_{\rm ave}(t)$ that would be obtained if the effective potential remained
time-independent. That is, we require that $\phi_{\rm ave}(t)$ oscillates in the effective
potential as determined immediately after the quench (i.e., without energy loss), just as the
parametric driving force in generalized Mathieu equations oscillates with constant amplitude. In
this sense, we are considering the initial response of the fluctuations to the oscillations of
$\phi_{\rm ave}(t)$, without accounting for the subsequent transfer of energy between the two.
Only with this approach can one achieve a constant exponential amplification of
$\bar{\delta\phi}$ and thus a good measure of resonance. Even though this approximation only
strictly holds for early times, the results obtained will confirm its validity in describing the
essential physics behind the observed synchronous emergence.

\begin{figure} \includegraphics[width=3in,height=3in]{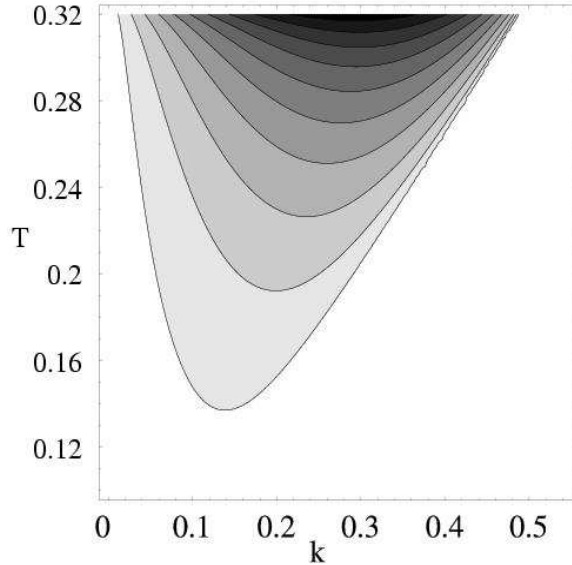} \caption {Lines of constant
amplification rate for small-amplitude modes at various temperatures, beginning with $\eta_{\rm
min}=2.8\times 10^{-2}$ for the bottom-most contour and increasing in increments of
$\Delta\eta=1.3\times 10^{-2}$.} \label{parares} \end{figure}

Figure \ref{parares} shows lines of constant exponential amplification rate $\eta$ of the 
fluctuations  $\delta\phi(k,t)$ for various $k$ and $T$. At low temperatures, $T<0.14$, no modes
are  ever amplified.  As the temperature is increased, so is the amplitude of oscillation in 
$\phi_{\rm ave}(t)$, gradually causing the band $0< k\leq 0.48$ to resonate. It is helpful at
this point to consider what modes comprise a typical oscillon. To do so, we use  a Gaussian
ansatz for the oscillon with a fixed radius  $R$, $\phi_{\rm osc}(r,t)\sim \exp(-r^2/R^2)$
\cite{gleiser}. The Fourier transform is itself a Gaussian with a characteristic radius $K=2/R$.
Thus, a Gaussian in real space is generally comprised of the band $0\leq k\leq 2/R$ in
$k$-space. Earlier [cf. figure \ref{dist}] it was found that the average radius of the emergent 
oscillons is $R_{\rm ave}=4.3$ at an initial temperature $T=0.22$, which in this approximation 
corresponds to the band $0\leq k\leq 0.47$. This agrees very well with what is seen in  figure
\ref{parares}. Although in this linear approximation the entire band does not resonate until
higher temperatures, it is expected that in the full equation of motion the nonlinear coupling
between modes will broaden the resonance band at lower temperatures. 

We have observed that the synchronous emergence of oscillons occurs for initial temperatures
$0.16\leq T\leq 0.26$. The lower bound is directly related to the nonlinear scattering between
modes, which is absent at low temperatures. For temperatures above $T\simeq 0.26$, the
appearance of large domains precludes the identification of oscillons. This is typical of the
approach to criticality, where large spatial correlations in the field become more probable,
destroying any local ordering. Thus, the observed quench-induced synchronous emergence of order
is limited to temperatures sufficiently below $T_c$.

\subsection{Oscillons as bottlenecks for equipartition} In the introduction, we mentioned that
long-lived locally-ordered structures may serve as bottlenecks to equipartition, as they slow
down the distribution of energy between the  system's various degrees of freedom. Below, we
present evidence that oscillons do precisely that. A more formal proof is left for future work.

A good observable to follow is the time evolution of $\langle\bar\pi^2(k,t)\rangle$ for
different modes,  which should become flat when thermal equilibrium is achieved and the
equipartition theorem is satisfied. In highly dissipative systems (such as those governed by a
Langevin equation of motion with a large viscosity coefficient), energy exchange between each
mode and the thermal bath tends to dominate the dynamics. In the nonequilibrium closed system
considered here, however, energy is only exchanged between the modes themselves, and the
thermalization process turns out to be quite different.

\begin{figure} \includegraphics[width=3in,height=3in]{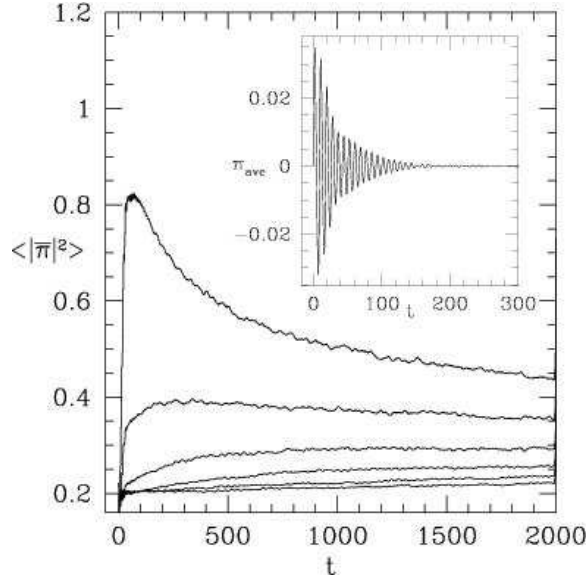} \caption {The time evolution
of $\langle\bar\pi^2(k,t)\rangle$ for various modes (see text). The data is  boxcar averaged
over a window $\Delta t=10$. Also shown (inset) is the time evolution of $\pi_{\rm ave}(t)$,
which attains equilibrium noticeably earlier than any of the higher modes shown.}
\label{thermal} \end{figure}

In figure \ref{thermal} we show the time evolution of $\langle\bar\pi^2(k,t)\rangle$ for the
values $k$=0.25, 0.5, 0.75, 1.0, 1.25, and 1.5 (top to bottom lines, respectively).  The curves
are time-averaged over a window $\Delta t=10$ so that the large oscillatory behavior at early
times ($t<60$) is suppressed. Also  shown in the inset is the time evolution of $\pi_{\rm
ave}(t)$, which  exhibits the thermalization trend of the zero mode. $\pi_{\rm ave}(t)$ displays
damped oscillatory motion before reaching its equilibrium value at $t\simeq 160$. This is
roughly the time  in which the mode $k=0.25$ reaches its maximum average kinetic energy. The
mode $k=0.5$ also increases from $t=0$ to a maximum value at about the same time before
decaying. For modes $k>0.5$, however, the general trend is to slowly absorb energy as they
asymptotically approach their equilibrium values from below on time-scales that can be much {\it
longer} than the lower-$k$ modes. There are two clear trends: small modes (with $k \leq 0.5$)
grow very quickly and then relax to their final equilibrium values, while larger modes (with $k
> 0.5$) slowly approach theirs. Thus, it is reasonable to conclude that most of the zero-mode's
initial energy goes into its neighboring modes, those responsible for oscillon-like
configurations [cf. figure \ref{parares}],  and only later on it is shared amongst larger $k$
modes. This seems to suggest that there is a {\it two-step energy cascade}, a fast one at early
times to small-$k$ modes, followed by a slower one to larger $k$ modes. The early emergence of
oscillons is consistent with this scenario.

\begin{figure} \includegraphics[width=3in,height=3in]{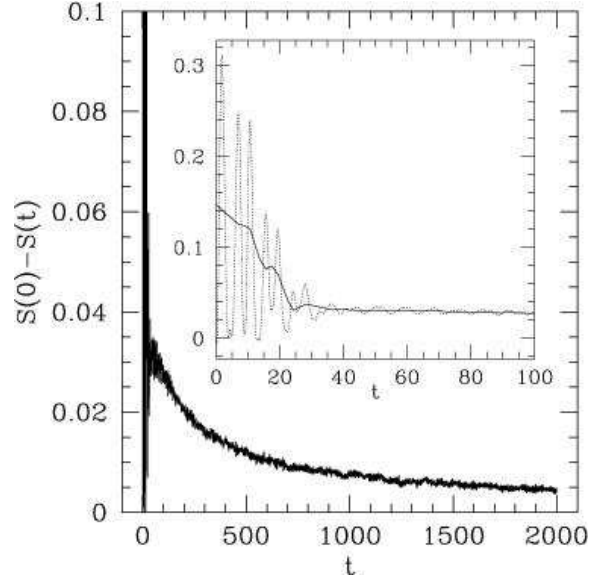} \caption {The change of $S(t)$
from the initial state for closed systems at $T=0.22$. The exponential approach to equilibrium
is clear at late times. The inset illustrates the role of  oscillons as a bottleneck to
equipartition.} \label{S} \end{figure}

We can further quantify this argument by introducing a measure of the entropy of the system
based on the partitioning of the kinetic energy,  $K({\bf k},t)=1/2|\bar\pi({\bf k},t)|^2$:
\begin{equation} S(t)=-\int_{A} d\,^2 k\,p({\bf k},t)\ln p({\bf k},t), \end{equation} where 
\begin{equation} p({\bf k},t)=\frac{K({\bf k},t)}{\int d\,^2 k\,K({\bf k},t)} \end{equation} is
the probability distribution of the kinetic energy in $k$-space. $S(t)$ attains its  maximum
value when equipartition is satisfied, and $p({\bf k},t)=A^{-1}$ where $A$ is the area of this
space.  This occurs both at the initial  thermalization $(t=0)$ and final equilibrium states,
since in these cases all modes carry the same  fractional kinetic energy. At these times we
recover the thermodynamic entropy of the system, $S_{\rm eq}=\ln A$, which is a measure of the
number of accessible  states in the system. [The numerical implementation gives $S_{\rm eq}=\ln
N$, where $N$ is the number of lattice points, or degrees of freedom.] In figure \ref{S} we show
the change of $S(t)$ from the  initial state, $S(t=0)-S(t)$, for the system at $T=0.22$. At late
times ($t>100$), we have found  that the system equilibrates exponentially in a time-scale $\tau
\simeq 500$. At early times, the  localization of energy at lower $k$ modes, corresponding to
the synchronous emergence of  oscillons, prolongs this  approach to equipartition. The inset of
figure \ref{S} shows the large oscillations in the early evolution of $S(t)$ (dotted  line). 
Also shown (solid line) is the average between successive peaks of $S(t)$, with a plateau at 
$30\leq t\leq 100$ that coincides with the maximum oscillon presence in the system.  This is to
be contrasted with an exponential approach to equilibrium. Thus, we see evidence that oscillon
configurations indeed serve as early bottlenecks to equipartition,  temporarily suppressing the
cascading of energy from low-$k$ modes to higher modes.

\section{Resonant Nucleation}

We will now investigate the dynamics with the ADW potential, that is, for values of $\alpha >
3/2$. Homogeneous nucleation predicts that, had the system been prepared in the metastable state
at $\phi=0$ at some temperature $T$, its decay rate per unit area would be controlled by the
Arrhenius factor, $\Gamma_{\rm HN} \sim \exp[-E_b/T]$, where $E_b$ is the free-energy barrier
for the nucleation of a critical nucleus or bounce \cite{langer,domb}. (We set $k_B=1$.) It is
often overlooked that this result is quite sensitive to how the initial state is prepared. It
assumes that the bounce is nucleated within a homogeneous background, by adopting a perturbative
Gaussian approximation for the evaluation  of the partition function. In other words, it assumes
that no large-amplitude perturbations are present in the system \cite{gleiser-heckler,tetradis}.
It is also well-known that there are discrepancies between the theoretical prediction from
homogeneous nucleation and a large array of experimental results concerning nucleation rates,
from He$^3$ \cite{leggett} and nematic liquid crystals \cite{degennes} to martensitic materials
and polycrystals \cite{offerman} to fluids \cite{bartell} and numerical simulations
\cite{gleiser_nuc}. More often than not, theoretical estimates greatly underestimate nucleation
rates. Surely, many of these discrepancies are related to limited knowledge of the
nonequilibrium processes leading to nucleation at the atomic or mesoscopic scale for various
materials. However, we propose here that in certain cases the problem is rooted on the detailed
set up of the initial metastable state. In particular, the quench we propose in this work, which
may be mimicked in experimental situations, generates an instability capable of greatly
accelerating the decay of the metastable state. In fact, we have observe that, for a range of
temperatures, the nucleation rate per unit area is not that of homogeneous nucleation,
$\Gamma_{\rm HN} = C\exp[-E_b/T]$, but a power law, $\Gamma_{\rm RN} = C (E_b/T)^{-B}$, where
$E_b(\alpha)$ is the critical nucleus free energy and the exponent $B$ is weakly
temperature-dependent.  RN stands for Resonant Nucleation. The power law applies precisely
within the temperature range where we observe the synchronous emergence of oscillons at early
times, $0.18 \leq T \leq 0.24$, as we show next.

The change in $\alpha$ will not modify the basic parametric amplification mechanism discussed
above in the context of the SDW: for a range of temperatures, oscillons are still nucleated in
synchrony at early times. The difference is that, for $\alpha > \alpha_c$,  the system must
reach the global  free-energy minimum at $\phi_+(\alpha)$. [Of course, for large enough
temperatures, $T>0.26$, and/or small energy barriers the system will go straight into the global
minimum, a cross-over decay. We are not interested in these cases.] 

\begin{figure} \includegraphics[width=3in,height=3in]{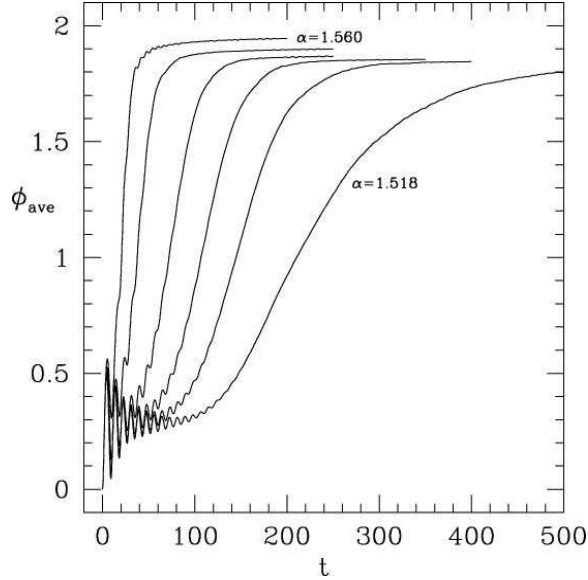} \caption {The evolution
of the order parameter $\phi_{\rm ave}(t)$  at $T=0.22$ for several values of the asymmetry.
(From left to right, $\alpha= 1.56, 1.542, 1.53, 1.524, 1.521, 1.518$.) Each curve is an ensemble
average over 100 runs.} \label{decay-phi} \end{figure}

In the usual treatment, the critical nucleation is described as the saddle-point in
field-configuration space, the unstable direction dictated by the negative eigenvalue associated
with the growth of the critical nucleus. This is given by the solution of the equation $\phi''
+(1/r)\phi' = V'$, with appropriate boundary conditions, which determines the 2d bounce
\cite{langer}. In the absence of forcing, which is the case for homogeneous nucleation, the
system will randomly probe phase space until it eventually ``finds'' the saddle point. The
presence of quench-induced large-amplitude fluctuations in the field will drastically accelerate
its decay. In figure \ref{decay-phi} we show the evolution of the order parameter $\phi_{\rm
ave}(t)$  as a function of time for several values of asymmetry, $1.518 \leq \alpha \leq 1.56$
for $T=0.22$. Not surprisingly, as $\alpha\rightarrow \alpha_c=1.5$, the field remains longer in
the metastable state, since the  nucleation energy barrier $E_b\rightarrow \infty$ at
$\alpha_c$. However, a quick glance at the time axis shows the fast decay time-scale, of order
$10^{1-2}$.  For comparison, for $1.52\leq \alpha \leq 1.56$, homogeneous nucleation would
predict  nucleation time-scales of order $\sim 10^{56} \leq \tau_{\rm HN}\leq 10^{21}$ (in
dimensionless units),  respectively. [The nucleation barriers are, $E_b(\alpha=1.158)=33.67$ and
$E_b(\alpha=1.56)=13.10$.] Clearly, while for smaller asymmetries $\phi_{\rm ave}(t)$ displays
similar oscillatory behavior to the  SDW case before transitioning to the global minimum, as
$\alpha$ is increased (asymmetry in potential is increased) the number of oscillations decreases
sharply. 

\begin{figure} \includegraphics[width=3in,height=3in]{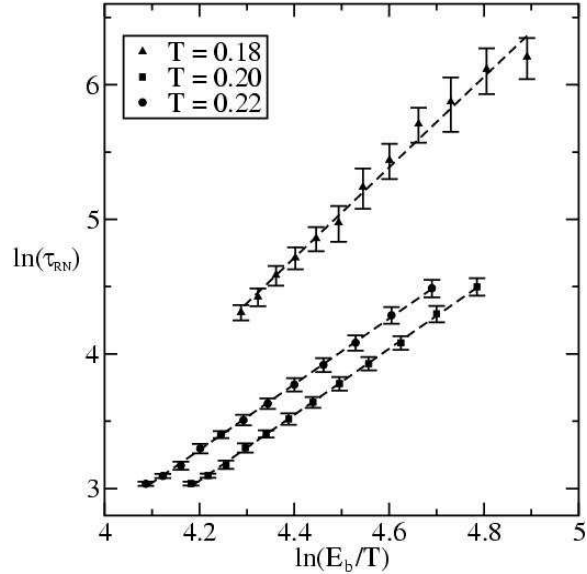} \caption {The decay
time-scale $\tau_{\rm RN}$ for resonant nucleation as a function of critical nucleation 
free-energy barrier $E_b/T$ at $T=0.18$, $0.2$, and $0.22$.  The best fit is a power-law with
exponent $B\simeq 2.44$ for $T=0.20$ and $0.22$ and $B=3.36$  for $T=0.18$.} \label{powerlaw}
\end{figure}

In figure \ref{powerlaw} we show the log-log plot of the ensemble-averaged nucleation
time-scales as a function of the nucleation barrier, or critical droplet free energy, $E_b/T$
for different temperatures, $T=0.18, 0.20,$  and $0.22$.  The error bars are computed from  the
dispersion of the measured time scales within the ensemble. The best fit is a power law,
$\tau_{RN} \propto (E_b/T)^B$, with $B=2.440 \pm 0.008$ for $T=0.20$ and $T=0.22$, and $B=3.36
\pm 0.04$ for $T=0.18$.  This simple power law holds for the same range of temperatures where we
have observed the synchronous emergence of oscillons. [The range is similar for both SDW and
ADW.] It is not surprising that the exponent $B$ is larger for $T=0.18$, where the synchronous
emergence of oscillons is marginal: at lower temperatures, the oscillations of the field's zero
mode have lower amplitude and oscillons become rarer. In fact, as we remarked before, we have
not observed any for $T<0.16$ for the SDW. For lower temperatures, we should thus expect a
smooth transition into the exponential time-scales of homogeneous nucleation.  We intend to
analyse this transition and to obtain an analytical explanation for the power law in a
forthcoming publication. As a first step in this direction, we present below what we believe is
the mechanism by which the transition completes for different nucleation barriers.

\begin{figure} \includegraphics[width=3in,height=3in]{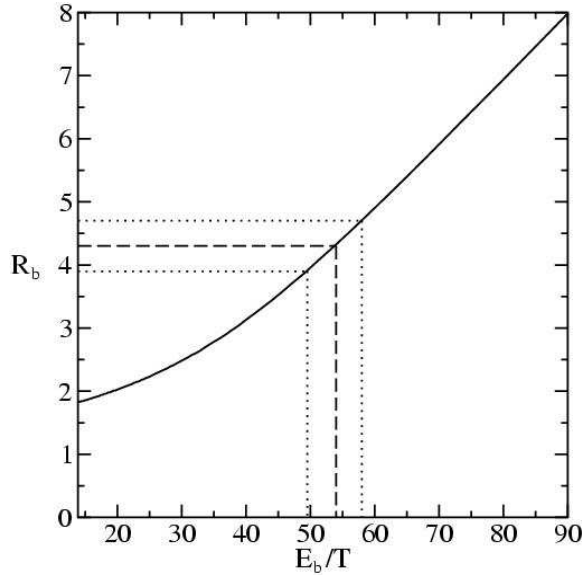} \caption {The critical
nucleus radius as a function of its energy barrier at $T=0.22$. The dashed line shows the
average radius for nucleated oscillons, while the dotted lines denote the dispersion of its
value.} \label{radius} \end{figure}

First, note that the error bars in figure \ref{powerlaw} increase with nucleation barrier. This
spread in time scales is related to the dynamics of the system. For $\alpha \rightarrow
\alpha_c$ the radius of the nucleation bubble diverges, $R_b \rightarrow \infty$. Clearly, due
to the disparity between the power law and the exponential time scales, the mechanism by which
the system approaches the global minimum is not through a random search in configuration space
as is the case for homogeneous nucleation. Instead, we argue that oscillons will act as seeds
for the nucleation of a critical fluctuation. The way in which  this happens depends on the
nucleation barrier.  If $\alpha \gtrsim \alpha_c$, the critical nucleus will have a radius much
larger than a typical oscillon [cf. figure \ref{dist}]; it will appear as two or more oscillons
coalesce. As $\alpha$ is increased further, the radius of the critical nucleus decreases,
approaching that of an oscillon. In this case, a single oscillon may become the critical nucleus
and promote the decay of the metastable state. This explains the small number of oscillations on
$\phi_{\rm ave}(t)$ as $\alpha$ is increased [cf. figure \ref{decay-phi}]. To corroborate our
argument, in figure \ref{radius} we contrast the critical nucleation radius with that of
oscillons for different values of energy barrier at $T=0.22$.  The average oscillon radius [cf.
figure \ref{dist}] is $R_{\rm ave} = 4.30 \pm 0.4$, and is identified by the dashed line, with
the dotted lines denoting its dispersion. The critical nucleus radius, $R_b$, is equal to
$R_{\rm ave}$ for $\alpha\simeq 1.575\pm 0.0075$, or $E_b/T=53.9\pm 0.4$. Thus,  for $\alpha
\gtrsim 1.568$ one cannot distinguish between a single oscillon and a critical bubble. Since, as
we have seen, oscillons appear very early on, the decay happens quite fast. For $1.5 < \alpha
< 1.568$,  approximately, the critical nucleus is the result of the coalescence of two or
more oscillons. They diffuse through the lattice and scatter, forming bound states, somewhat as
in kink-antikink breathers in 1d field theory \cite{campbell,hanggi}.  We emphasize that these
values are only approximate; they do not include the increase in coalescence rates
due to diffusive  motions on the lattice or an enhancement on the
unstable growth of oscillons due to 
perturbations. We expect that the one-oscillon decay will ensue for smaller values of
$\alpha$. In \cite{website} the reader can see a few representative simulations.

Clearly, these arguments can be refined. In particular, the details of oscillon coalescence
remain unexplored. However, they encapsulate the basic mechanism by which metastable states
decay via resonant nucleation, as seen in numerical simulations.

\section{Summary and Outlook}

We have investigated the nonequilibrium dynamics of a double-well model with a 2d scalar order
parameter. The system was initially prepared in thermal equilibrium at a fixed temperature in a
anharmonic single-well potential, $V(\phi)=\frac{1}{2}\phi^2 +\frac{1}{8}\phi^4.$  We then
turned on a cubic interaction with tunable strength so that the resulting double-well potential
can be  symmetric (SDW) or asymmetric (ADW). The turning-on of the interaction can be
interpreted as a ``quench,''  as it occurs within a time-scale faster than any in the system.
The system is thus tossed into a double-well potential and, through its nonlinear interactions,
redistributes its energy until it reaches equipartition. For the SDW case, we have observed that
this approach to equilibrium occurs as a two-step energy cascade: quickly at first, as the
zero-mode relinquishes its energy to nearby (in $k$-space) long-wavelength modes, and  slower
later, as the energy is partitioned with the short-wavelength modes. Although we have adopted
the language of nonequilibrium  field theory, we remark that one could equally well treat the
system as a microcanonical ensemble with fixed energy $E$. 

We showed that, for a range of initial temperatures, the initial stages of the dynamics are
dominated by damped oscillations of the field's zero mode, which, via parametric resonance,
excite the modes associated with coherent structures named oscillons. We also showed that the
initial emergence of oscillons occurs at the breakdown of  the homogeneous Hartree
approximation. Furthermore, this emergence occurs in synchrony, characterizing a  global
emergence phenomenon. As oscillons are long-lived coherent structures, they act as bottlenecks
for  equipartition. The observed synchronous emergence dies away with the damping of the
zero-mode oscillations.  Oscillons then appear randomly throughout the lattice, until such a
time when there isn't enough energy in  long-wavelength modes.

We then investigated the dynamics in ADWs, where the final equilibrium state is at the global
minimum of the potential. The system was thermalized as for the SDW, in its metastable well.
Also as before, the quenching induces damped oscillations of the field's zero mode and oscillons
emerge. We observed that, approximately for the same range of temperatures where oscillons
emerged in the SDW case, the decay rate was greatly enhanced, when compared to homogeneous
nucleation results. In fact, we obtained an excellent fit to a {\it power law decay} time scale,
$\tau_{\rm RN} \propto [E_b/T]^B$, where $E_b$ is the energy  of the critical nucleus or bounce
and $B$ is a numerical exponent, $2.44 \leq B\leq 3.36$, for the temperatures  we investigated.
We argued that there were two main mechanisms for what we called resonant nucleation: for small
asymmetries, the critical nucleus appears as two or more oscillons coalesce. For larger
asymmetries, a single oscillon grows to become the critical nucleus, promoting the decay of the
metastable state. 

The greatly accelerated decay is due to the resonant oscillations of the zero mode, induced by
the quenching. It is thus plausible that such behavior may be observed in several related
systems with a scalar order parameter. The emergence of oscillons in granular materials appears
due to the sinusoidal vibrations of their container \cite{umbanhowar};  this is mimicked here by
the oscillation of the field's zero mode, which corresponds to the periodic oscillation of the
whole lattice. As for granular materials, patterns emerge above a certain amplitude. We do not
claim to have modelled the behavior of granular materials here, but believe to have captured
some of the essential physics. Also, since the model here falls in the Ising universality class,
we expect that similar qualitative behavior should occur, for example, in ferromagnets and
possibly binary liquids and metal alloys, if long wavelength oscillations can be induced by the
quenching process.

\acknowledgments

MG was support in part by a National Science foundation grant PHY-0099543. We thank J. D. Gunton
and J. A. Krumhansl for their comments and insights.


\begin{thebibliography}{99}

\bibitem{walgraef}D. Walgraef, {\it Spatio-temporal Pattern Formation} (Springer, New York,
1997); A. I. Olemskoi and V. F. Klepikov, Phys. Rep. {\bf 338}, 571 (2000).

\bibitem{bar-yam}Y. Bar-Yam, {\it Dynamics of Complex Systems} (Addison-Wesley, Reading, MA,
1997).

\bibitem{cross}M. C. Cross and P. C. Hohenberg, Rev. Mod. Phys. {\bf 65}, 851 (1993); Y.
Kuramoto, {\it Chemical Oscillations, Waves, and Turbulence} (Springer, Berlin, 1984).

\bibitem{phase}N. Goldenfeld, {\it Lectures on Phase Transitions and The Renormalization 
Group}, Frontiers in Physics, V. 85 (Addison-Wesley, NY, 1992).

\bibitem{magnets}A. Aharoni, {\it Introduction to the Physics of  Ferromagnetism} (Oxford
University Press, NY, 1996).

\bibitem{degennes}P. G. de Gennes, {\it The Physics of Liquid Crystals} (Oxford University
Press, Oxford, 1993).

\bibitem{schnerb}N. M. Shnerb, Y. Louzoun, E. Bettelheim, and S. Solomon, S., PNAS, {\bf 97},
10322-10324 (2000).

\bibitem{ford}J. Ford, Phys. Rep. {\bf 213}, 271 (1992).

\bibitem{discrete}S. Flach and C. R. Willis, Phys. Rep. {\bf 295}, 181 (1998)

\bibitem{bottlenecks}R. Reigada, A. Sarmiento, and K. Lindenberg, cond-mat/0210326

\bibitem{rajaraman}R. Rajamaran, {\it Solitons and Instantons} (North-Holland,  Amsterdam,
1987); T. D. Lee and Y. Pang,  Phys. Rep. {\bf 221}, 251 (1992).

\bibitem{vilenkin}A. Vilenkin and E. P. S. Shellard, {\it Cosmic Strings and Other Topological
Defects} (Cambridge University Press, Cambridge, 1994).

\bibitem{NTSs}S. Coleman, {\it Aspects of Symmetry}, Cambridge University Press (Cambridge, UK
1985)

\bibitem{campbell}D. K. Campbell, J. F. Schonfeld, and C. A. Wingate, Physica {\bf 9D}, 1
(1983).

\bibitem{digal}For an exception see, S. Digal, R. Ray, S. Sengupta, and A. M. Srivastava, Phys.
Rev. Lett. {\bf 84}, 826 (2000).

\bibitem{gleiser-howell}M. Gleiser and R. Howell, arXiv:cond-mat/0209176, in press Phys. Rev. E
(RC).

\bibitem{gleiser}M. Gleiser, Phys. Rev. D {\bf 49}, 2978 (1994);  E. J. Copeland, M. Gleiser,
and H. R. Muller, Phys. Rev. D {\bf 52}, 1920 (1995); E. B. Bogomol'nyi, Sov. J. Nucl. Phys.
{\bf 24}, 449 (1976).

\bibitem{bettinson}D. Bettison and G. Rowlands, Phys. Rev. E {\bf 55}, 5427 (1997); C. Crawford
and H. Riecke, Phys. Rev. E {\bf 65}, 066307 (2002).

\bibitem{umbanhowar}P. Umbanhowar, F. Melo, and H. Swinney, Nature {\bf 382}, 793 (1996); L. S.
Tsimring and I. S. Aranson, Phys. Rev. Lett. {\bf 79}, 213 (1997); S.-O. Jeong and H.-T. Moon,
Phys. Rev. E {\bf 59}, 850 (1999).

\bibitem{umurhan}O. M. Umurhan, L. Tao, and E. A. Spiegel, Ann. N. Y. Acad. Sci. {\bf 867},  298
(1998).

\bibitem{langer}J. S. Langer, in {\it Solids Far from Equilibrium}, Ed. C. Godr\`eche (Cambridge
University Press, Cambridge, 1992).

\bibitem{domb}J. D. Gunton, M. San Miguel, and P. S. Sahni, in {\it Phase Transitions and
Critical Phenomena}, Ed. C. Domb and J. L. Lebowitz, v. 8 (Academic Press, London, 1983); J. D.
Gunton, J. Stat. Phys. {\bf 95}, 903 (1999).

\bibitem{howell}M. Gleiser, R. C. Howell, and R. O. Ramos, Phys. Rev. E {\bf 63}, 036113 (2002).

\bibitem{munoz}M. A. Mu\~noz and R. Pastor-Satorras, Phys. Rev. Lett. {\bf 90}, 204101 (2003).

\bibitem{chandler}D. Chandler, {\it Introduction to Modern Statistical Mechanics}, (Oxford
University Press, Oxford, 1987).

\bibitem{bonini}G. Aarts, G. F. Bonini, and C. Wetterich, Phys. Rev. D {\bf 63}, 025012 (2000);
G. Aarts, G. F. Bonini, and C. Wetterich, Nucl. Phys. B {\bf 587}, 403 (2000).

\bibitem{boyanovsky:03}D. Boyanovsky, C. Destri, and H. J. de Vega, arXiv:hep-ph/0306124.

\bibitem{borrill}J. Borrill and M. Gleiser, Nucl. Phys. B {\bf 483}, 416 (1997).

\bibitem{gleiser-mix}M. Gleiser, Phys. Rev. Lett. {\bf 73}, 3495 (1994); J. Borrill and M.
Gleiser, Phys. Rev. {\bf D51}, 4111 (1995).

\bibitem{website}see http://www.dartmouth.edu/$\sim$cosmos/oscillons

\bibitem{numericalrecipes}W. Press, S. Teukolsky, W. Vetterling, and B. Flannery, {\it Numerical
Recipes in C}, (Cambridge University Press, Cambridge UK, 1992).

\bibitem{sornborger}M. Gleiser and A. Sornborger, Phys. Rev. E {\bf 62}, 1368 (2000);A. Adib, M.
Gleiser, and C. Almeida, Phys. Rev. D {\bf 66}, 085011 (2002).

\bibitem{kofman}Here is an incomplete list of references:  G. N. Felder and L. Kofman, Phys.
Rev. D {\bf 63}, 103503 (2001); P. B. Greene and L. Kofman, Phys. Lett. B {\bf 448}, 6 (1999);
P. B. Greene, L. Kofman, A. D. Linde and A. A. Starobinsky, Phys. Rev. D {\bf 56}, 6175 (1997);
D. Boyanovsky, H. J. de Vega and R. Holman, arXiv:hep-ph/9701304; D. Boyanovsky, M. D'Attanasio,
H. J.de Vega, R. Holman and D. S. Lee, Phys. Rev. D {\bf 52}, 6805 (1995); D. Boyanovsky, H. J.
de Vega and R.~Holman, arXiv:hep-ph/9903534.

\bibitem{gleiser-heckler}M. Gleiser and A. Heckler, Phys. Rev. Lett. {\bf 76}, 180 (1996).

\bibitem{tetradis}A. Strumia, N. Tetradis and C. Wetterich, Phys. Lett. B {\bf 467}, 279
(1999). 

\bibitem{leggett}A. J. Leggett and S. K. Yip, in {\it Helium Three}, edited by W. P. Halperin
and L. P. Pitaevskii (North-Holland, Amsterdam, 1990).

\bibitem{offerman} S. E. Offerman {\it et al.}, Science {\bf 298} (2002) 1003.

\bibitem{bartell}J. Huang and L. S. Bartell, J. Phys. Chem. {\bf 99}, 3924 (1994).

\bibitem{gleiser_nuc}M. Alford and M. Gleiser, Phys. Rev. {\bf D48}, 2838 (1993); G. T. Valls
and G. F. Mazenko, Phys. Rev. B {\bf 42}, 6614 (1990);   K. Park, P. A. Rikvold, G. M. Buendia,
M.A. Novotny, arXiv:cond-mat/0307595.

\bibitem{hanggi} P. H\"anggi, F. Marchesoni, and P. Sodano, Phys. Rev. Lett. {\bf 60}, 2563
(1988).

\end{thebibliography}
\end{document}